\documentclass[twocolumn,preprintnumbers,amsmath,amssymb,amsfonts,floatfix,nofootinbib]{revtex4}
\usepackage{graphicx}
\usepackage{hypernat}
\usepackage{color}
\usepackage{ifpdf}
\usepackage{bm}
\usepackage{dcolumn}

\newif\ifcomment
\newif\ifprint
\printtrue

\ifpdf
\ifprint
 \usepackage[linkbordercolor={0 0 .8},%
             citebordercolor={.8 0 0},%
             urlbordercolor={.8 0 .8},%
             raiselinks=true,%
             pdfborder={0 0 1 [3]}]{hyperref}
\else
 \usepackage[pdftex,backref,pagebackref,colorlinks,
             linkcolor=blue,citecolor=blue,anchorcolor=blue,
             pagecolor=blue,urlcolor=red]{hyperref}
\fi
\pdfoutput=1        
\pdfcatalog{/PageMode(/UseOutlines)} 
\else 
\ifprint
 \usepackage[linkbordercolor={0 0 .8},%
             citebordercolor={.8 0 0},%
             urlbordercolor={.8 0 .8},%
             raiselinks=true,%
             pdfborder={0 0 1 [3]}]{hyperref}
\else
 \usepackage[dvips,backref,pagebackref,colorlinks,
             linkcolor=blue,citecolor=blue,anchorcolor=blue,
             pagecolor=blue,urlcolor=red]{hyperref}
\fi
\hypersetup{
  pdfauthor   = {},
  pdftitle    = {},
  pdfcreator  = {},
  pdfproducer = {},
  pdfsubject  = {},
  pdfkeywords = {}
 }
\fi

\graphicspath{{./figures/}}

\def\myepsilon{\epsilon}
\def\MCG{MCG}

\newcommand {\av}[1]       {\ensuremath{\left \langle #1 \right \rangle}}
\newcommand {\avsm}[1]     {\ensuremath{\langle {#1} \rangle}}
\newcommand {\abs}[1]      {\ensuremath{\left| #1 \right|}}
\newcommand {\ordof}[1]    {\ensuremath{{\cal O}\left( {#1} \right)}}
\newcommand {\snn}         {\ensuremath{\sqrt{s_{\scriptscriptstyle{{\rm NN}}}}}}
\newcommand {\signn}       {\ensuremath{\sigma_{\scriptscriptstyle{{\rm NN}}}}}
\newcommand {\es}          {\ensuremath{\myepsilon_{\rm s}}}
\newcommand {\ep}          {\mbox{$\myepsilon_{\rm part}$}}
\newcommand {\eptwo}       {\mbox{$\myepsilon_{\rm part}\{2\}$}}
\newcommand {\epfour}      {\mbox{$\myepsilon_{\rm part}\{4\}$}}
\newcommand {\epsq}        {\mbox{$\myepsilon^2_{\rm part}$}}
\newcommand {\epfr}        {\mbox{$\myepsilon^4_{\rm part}$}}
\newcommand {\erp}         {\mbox{$\myepsilon_{\rm RP}$}}

\newcommand {\epart}       {\ep}
\newcommand {\spart}       {\ensuremath{\sigma_{\myepsilon_{\rm part}}}}
\newcommand {\psipart}     {\ensuremath{\Psi_{\rm part}}}
\newcommand {\vrp}         {\ensuremath{v_{2}\{{\rm EP}\}}}

\newcommand {\epsrp}       {\ensuremath{\myepsilon\{{\rm EP}\}}}

\newcommand {\Ncoll}       {\ensuremath{N_{\rm coll}}}
\newcommand {\Npart}       {\ensuremath{N_{\rm part}}}

\newcommand {\fbkg}         {\ensuremath{f_{\rm bkg}}}

\newcommand {\dncdy}       {\ensuremath{\mathrm{d}N_{\mathrm{ch}}/\mathrm{d}y}}
\newcommand {\dncde}       {\ensuremath{\mathrm{d}N_{\mathrm{ch}}/\mathrm{d}\eta}}
\newcommand {\dnppcde}     {\ensuremath{\mathrm{d}N^{\rm pp}_{\mathrm{ch}}/\mathrm{d}\eta}}

\newcommand {\pt}          {\ensuremath{p_{\mathrm{T}}}}

\newcommand {\AuAu}        {${\rm Au} + {\rm Au}$}
\newcommand {\CuCu}        {${\rm Cu} + {\rm Cu}$}

\newcommand {\rsq}         {\mbox{$\langle r^2 \rangle$}}

\newcommand {\boden}[2]    {\ensuremath{\frac{{#1}}{\rsq^{{#2}}}}}
\newcommand {\dysqp}       {\ensuremath{\delta_{r^2}}}
\newcommand {\dysqm}       {\ensuremath{\delta_{y^2-x^2}}}
\newcommand {\dsqyp}       {\ensuremath{(\delta^2_{y}+\delta^2_{x})}}
\newcommand {\dsqym}       {\ensuremath{(\delta^2_{y}-\delta^2_{x})}}

\newcommand {\dsqysqm}     {\ensuremath{\delta_{r^2}\delta_{y^2-x^2}}}
\newcommand {\dd}          {\mathrm{d}}

\newcommand {\lsim}        {\,{\buildrel < \over {_\sim}}\,}

\newcommand {\Ref}[1]      {Ref.~\cite{#1}}
\newcommand {\Refs}[1]     {Refs.~\cite{#1}}
\newcommand {\Eq}[1]       {Eq.~(\ref{#1})}
\newcommand {\Equa}[1]     {Equation~(\ref{#1})}
\newcommand {\Eqn}[1]      {Eq.~({#1})}

\newcommand {\Sect}[1]     {Section~\ref{#1}}
\newcommand {\App}[1]      {Appendix~\ref{#1}}

\newcommand {\Fig}[1]      {Fig.~\ref{#1}}
\newcommand {\Tab}[1]      {Table~\ref{#1}}

\newcommand {\hide}[1]     {\color{white}#1\color{black}}

\newcommand {\hrefurl}[1]  {\href{#1}{\url{#1}}}
\newcommand {\eg}          {e.g.}
\newcommand {\ie}          {i.e.}

\newcommand {\etc}         {etc.}
\newcommand {\wrt}         {with respect to}
\newcommand {\beq}         {\begin{equation}}
\newcommand {\eeq}         {\end{equation}}
\newcommand {\beqa}        {\begin{eqnarray}}
\newcommand {\eeqa}        {\end{eqnarray}}
\newcommand {\beqnn}       {\begin{equation*}}
\newcommand {\eeqnn}       {\end{equation*}}
\newcommand {\beqann}      {\begin{eqnarray*}}
\newcommand {\eeqann}      {\end{eqnarray*}}

\begin{document}

\title{The Importance of Correlations and Fluctuations\\ 
on the Initial Source Eccentricity in High-Energy Nucleus--Nucleus Collisions}

\author{
B.Alver$^4$,
B.B.Back$^1$,
M.D.Baker$^2$,
M.Ballintijn$^4$,
D.S.Barton$^2$,
R.R.Betts$^7$,
R.Bindel$^8$,
W.Busza$^4$,
V.Chetluru$^7$,
E.Garc\'{\i}a$^7$,
T.Gburek$^3$,
J.Hamblen$^9$,
U.Heinz$^6$,
D.J.Hofman$^7$,
R.S.Hollis$^7$,
A.Iordanova$^7$,
W.Li$^4$,
C.Loizides$^4$,
S.Manly$^9$,
A.C.Mignerey$^8$,
R.Nouicer$^2$,
A.Olszewski$^3$,
C.Reed$^4$,
C.Roland$^4$,
G.Roland$^4$,
J.Sagerer$^7$,
P.Steinberg$^2$,
G.S.F.Stephans$^4$,
M.B.Tonjes$^8$,
A.Trzupek$^3$,
G.J.van~Nieuwenhuizen$^4$,
S.S.Vaurynovich$^4$,
R.Verdier$^4$,
G.I.Veres$^4$,
P.Walters$^9$,
E.Wenger$^4$,
B.Wosiek$^3$,
K.Wo\'{z}niak$^3$,
B.Wys\l ouch$^4$\\
\vspace{3mm}
\small
%
%
 $^1$~Physics Division, Argonne National Laboratory, Argonne, IL 60439-4843, USA\\
 $^2$~Physics and C-A Departments, Brookhaven National Laboratory, Upton, NY 11973-5000, USA\\
 $^3$~Institute of Nuclear Physics, Krak\'{o}w, Poland\\
 $^4$~Laboratory for Nuclear Science, Massachusetts Institute of Technology, Cambridge, MA 02139-4307, USA\\
 $^5$~Department of Physics, National Central University, Chung-Li, Taiwan\\
 $^6$~Department of Physics, Ohio State University, Columbus, OH 43210, USA\\
 $^7$~Department of Physics, University of Illinois at Chicago, Chicago, IL 60607-7059, USA\\
 $^8$~Department of Chemistry, University of Maryland, College Park, MD 20742, USA\\
 $^9$~Department of Physics and Astronomy, University of Rochester, Rochester, NY 14627, USA\\
}

\date{\today,\hspace{0.2cm}$$Revision: 1.66 $$}

\begin{abstract}
In relativistic heavy-ion collisions, anisotropic collective flow is driven,
event by event, by the initial eccentricity of the matter created in the
nuclear overlap zone.
Interpretation of the anisotropic flow data thus requires a detailed
understanding of the effective initial source eccentricity of the event
sample. In this paper, we investigate various ways of defining this effective
eccentricity using the Monte Carlo Glauber~(\MCG) approach.
In particular, we examine the participant eccentricity, which quantifies the 
eccentricity of the initial source shape by the major axes of the ellipse 
formed by the interaction points of the participating nucleons. 
We show that reasonable variation of the density parameters in the Glauber calculation, 
as well as variations in how matter production is modeled, do not significantly modify 
the already established behavior of the participant eccentricity as a function of collision 
centrality.
Focusing on event-by-event fluctuations and correlations of the distributions of participating 
nucleons we demonstrate that, depending on the achieved event-plane resolution, fluctuations in 
the elliptic flow magnitude $v_2$ lead to most measurements being sensitive to the root-mean-square, 
rather than the mean of the $v_2$ distribution.
Neglecting correlations among participants, we derive analytical expressions for the participant 
eccentricity cumulants as a function of the number of participating nucleons, $\Npart$,
keeping non-negligible contributions up to~$\ordof{1/\Npart^3}$.
We find that the derived expressions yield the same results as obtained from mixed-event \MCG\ 
calculations which remove the correlations stemming from the nuclear collision process.
Most importantly, we conclude from the comparison with \MCG\ calculations that the fourth order 
participant eccentricity cumulant does not approach the spatial anisotropy obtained assuming a smooth 
nuclear matter distribution.
In particular, for the \CuCu\ system, these quantities deviate from each other by almost a factor 
of two over a wide range in centrality.
This deviation reflects the essential role of participant spatial correlations in the interaction 
of two nuclei. 
\end{abstract}


\maketitle

\section{\label{intro}Introduction}

One of the strongest pieces of evidence for the formation of a thermalized
dense state of unconventional strongly interacting matter in 
ultra-relativistic nucleus-nucleus collisions at the Relativistic Heavy 
Ion Collider~(RHIC)~\cite{Arsene:2004fa,Adcox:2004mh,Back:2004je,Adams:2005dq}
stems from the strong anisotropic collective flow measured in non-central
collision events~\cite{Adcox:2002ms,Adler:2003kt,Adler:2004cj,%
Back:2002gz,Back:2004mh,Back:2004zg,Ackermann:2000tr,Adams:2004bi}. 
Studies of the final charged particle momentum distributions have revealed 
strong collective effects in the form of anisotropies in the azimuthal distribution
transverse to the direction of the colliding nuclei, 
and theory holds that their anisotropy around the beam axis in 
non-central collisions is established during the earliest stages of
the evolution of the collision fireball~\cite{Sorge:1996pc,Sorge:1998mk,%
Kolb:2000sd,Heinz:2001xi}. The main component of this anisotropy is called 
``elliptic flow'' and its strength is commonly quantified by the second 
coefficient, $v_2$, in the Fourier decomposition of the azimuthal momentum 
distribution of observed particles relative to the reaction 
plane~\cite{Poskanzer:1998yz}. 

By now there exists an extensive data set of elliptic flow measurements
in \AuAu\ collisions at RHIC as a function of center-of-mass energy,
centrality, pseudo-rapidity and transverse momentum~\cite{Adcox:2002ms,
Adler:2003kt,Adler:2004cj,Back:2002gz,Back:2004mh,Back:2004zg,
Ackermann:2000tr,Adams:2004bi}. The magnitude of the observed flow 
anisotropy is found to be strongly correlated with the anisotropic 
shape of the initial nuclear overlap region. This is expected if 
interactions among the initially produced particles are very strong,
leading to anisotropic pressure gradients,
which transform the initial spatial eccentricity into a final momentum 
anisotropy~\cite{Ollitrault:1992bk}. 

Quantitatively, the connection between 
initial spatial and final momentum anisotropy is explored by hydrodynamical 
calculations that, for a given equation of state, relate a given initial 
source distribution to the final momentum distribution of the produced
particles. For \AuAu\ collisions at the top RHIC energy, \mbox{$\snn=200\,$GeV}, 
such calculations are in good agreement with the elliptic flow data at 
mid-rapidity~\cite{Kolb:2000fh,Back:2004je}. 
From similar studies, it has been numerically established that, for not
too large impact parameters, the final magnitude of the elliptic flow is 
proportional to the initial eccentricity, $\myepsilon$, used to characterize
the spatial anisotropy in the transverse plane of the matter created in the 
overlap region of the colliding nuclei~\cite{Kolb:2000sd,Bhalerao:2005mm}.
More generally, one expects the ratio of elliptic flow and eccentricity, 
$v_2/\myepsilon$, at mid-rapidity to be a universal function of density and 
size of the system at the time when the elliptic flow 
develops~(``$v_2/\myepsilon$ scaling'')~\cite{Heiselberg:1998es,Voloshin:1999gs,
Bhalerao:2006tp}. In hydrodynamics, this function depends parametrically on 
the speed of sound in the fireball medium~\cite{Bhalerao:2005mm}.

The elliptic flow in \CuCu\ collisions at RHIC was found to be 
comparatively large, especially for near-central collisions, reaching 
almost the same magnitude as in \AuAu\ collisions for the same fractional 
cross section~\cite{Alver:2006wh,Adare:2006ti}, and much larger than 
expected from hydrodynamical models~\cite{Hirano:2005hv}. 
A quantitatively meaningful comparison 
of the elliptic flow values measured in \CuCu\ and \AuAu\ collisions requires
dividing out the difference in the eccentricity of the nuclear overlap 
zone since, for a given centrality, the average eccentricity depends on 
the size of the colliding nuclei.
For the same size of the overlap zone similar densities are achieved
in the two collision systems~\cite{Alver:2005nb,Roland:2005ei}, but the 
\CuCu\ system exhibits a significantly smaller spatial eccentricity. If
one scales the measured $v_2$ by this eccentricity, using its conventional 
definition in terms of the spatial deformation of the average transverse 
distribution of participating nucleons at a given impact parameter,
one is led to the paradoxical finding that the smaller \CuCu\ system translates 
the initial spatial deformation more efficiently into a final momentum 
anisotropy than the larger \AuAu\ system~\cite{Alver:2005nb,Roland:2005ei}. 

\begin{figure}[t]
\includegraphics[bb=50 80 980 760, width=0.4\textwidth]{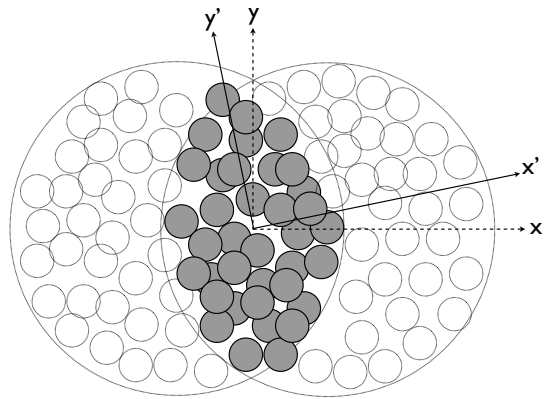}
\caption{\label{fig:eccpartscheme}
Schematic illustration of a nucleus--nucleus collision depicted in the transverse plane. 
The principal axes~($x^\prime$ and $y^\prime$) of the area formed by the 
participants are tilted \wrt\ the reaction plane
given by the axes~($x$ and $y$) of the transverse plane.}
\end{figure}

However, this conclusion depends on the definition of the eccentricity
used in the scaling procedure. 
In \Ref{Miller:2003kd} it was pointed out that the shape of the nuclear 
overlap fluctuates from event to event, and in \Ref{Manly:2005zy} it is
emphasized that the orientation of its major axes relative to the 
reaction plane~(defined by the beam axis and impact parameter vector) 
fluctuates as well. This is illustrated in \Fig{fig:eccpartscheme}.
For a given event, the actual distribution of the participant interaction 
points in the overlap zone can cause the overlap zone to be tilted \wrt\ 
the reaction plane.
The participant eccentricity~($\epart$, \Eq{eq:epspart}~\cite{Manly:2005zy})
takes this into account by using the principal axes of the overlap zone.
Because of the fluctuations, the ensemble average of the event-wise 
participant eccentricity
is not identical with the standard eccentricity~($\es$, \Eq{eq:epsstd}) 
of the smooth overlap distribution which is obtained by averaging the 
participant density in the overlap region \wrt\ the reaction plane over 
many events. 

Since hydrodynamic collective flow is not a property of the event 
ensemble, but rather develops independently {\em in each collision event}, 
its driving force is the shape and deformation of the initial distribution 
of produced matter {\em in each event}. To investigate the validity of the 
hydrodynamically predicted $v_2/\myepsilon$ scaling one should therefore, 
in principle, compute the ratio $v_2/\ep$ event by event, before taking 
its ensemble average $\av{v_2/\ep}$. This is, unfortunately, 
not possible in practise since the initial spatial eccentricity $\ep$ of 
a given collision event cannot be measured, and a statistically accurate 
determination of the elliptic flow $v_2$ also requires summing the hadron 
momentum spectra over many events, so only its ensemble average $\av{v_2}$ 
is known. In practise, the best way to approximate $\av{v_2/\ep}$
is to scale the measured elliptic flow $\av{v_2}$ by a calculated average 
participant eccentricity $\av{\ep}$ or by a higher moment of it~(see below).

In~\Ref{Manly:2005zy,Alver:2006wh} the participant eccentricity scaling was studied
using Monte Carlo Glauber~(\MCG) calculations, where $\ep$ is computed for each event 
from the transverse position distribution of nucleons participating 
in the collision, taken in its individual major axis frame.
For large nuclei, event-wise fluctuations in the 
transverse density distributions are small, except for the most peripheral 
collisions. Nonetheless, as one approaches zero impact parameter~(\ie~in 
almost central collisions where both $\es$ and $\ep$ are tiny), these 
small density fluctuations still cause significant relative fluctuations 
of $\ep$, resulting in a non-negligible difference between the participant
and the standard eccentricity, even in \AuAu\ collisions.
For the smaller \CuCu\ system the event-wise fluctuation 
effects are much stronger and seriously affect the eccentricity over 
the entire range of impact parameters~\cite{Alver:2006wh}. 
The participant-eccentricity-scaled elliptic flow $\av{v_2}/\av{\ep}$ thus differs 
appreciably from the standard eccentricity-scaled elliptic 
flow. It appears that scaling $v_2$ with the participant eccentricity unifies the 
eccentricity-scaled elliptic flow $\av{v_2}/\av{\epart}$ across 
the \CuCu\ and \AuAu\ collision systems~\cite{Manly:2005zy,Alver:2006wh}, 
even differentially as a function of transverse momentum and 
pseudo-rapidity~\cite{Nouicer:2007rs}. 
Furthermore, first measurements of elliptic flow fluctuations have recently been 
reported in \AuAu\ collisions at $\snn=200\,$GeV~\cite{Alver:2007qw,Sorensen:2006nw}.
The relative fluctuation magnitude $\spart/\avsm{\epart}$ from \MCG\ is in 
striking agreement with $\sigma_{v_2}/\avsm{v_2}$ from data~\cite{Alver:2007qw},
as expected if initial state fluctuations are combined with hydrodynamic 
evolution.

The initial success of the participant eccentricity calculated in the
\MCG\ approach immediately suggests a new set of questions:
\begin{itemize}
\item How robust are the participant eccentricity results to the parameters 
characterizing the nuclear density distribution~(radius, skin depth and 
nucleon--nucleon potential)?
\item What is the effect of varying the assumptions about matter 
production~(locality, participant and binary collision weighting)?
\item What features of the \MCG\ initial state distinguish it from the 
usual optical Glauber model picture? 
\item More specifically, what is the
impact of the fluctuating initial conditions on the 
suggested~\cite{Miller:2003kd,Bhalerao:2006tp} use of cumulant approaches?
In particular, which moment of an underlying fluctuating flow 
distribution is measured by the~(standard) event-plane flow 
method~\cite{Poskanzer:1998yz}?
\end{itemize}

These questions will be addressed in the present paper. In addition,
following and improving on \Ref{Bhalerao:2006tp}, we derive analytical expressions 
for the eccentricity cumulants in terms of moments of the initial spatial 
matter distribution, including all leading terms.
Furthermore, by comparing with the numerical \MCG\ model, we show that the analytical
expressions are misleading as they neglect important effects arising
from spatial correlations between the participating nucleons. 

\section{\label{montecarloglauber}Monte Carlo Glauber model}

To estimate the geometrical configurations of colliding nuclei, one typically constructs models based on rather generic 
assumptions about the constituent makeup of a  typical nucleus. In this context, it is fairly standard to assume that 
nuclear matter in a nucleus is distributed according to the charge distributions seen in electron scattering experiments.
There are two ways of expressing these densities in actual calculations~(for an overview, see \Ref{Miller:2007ri} and 
references therein). One way is to assume a smooth matter density, typically described by a Fermi distribution in 
the radial direction and uniform over solid angle~(in the case of spherical or near-spherical nuclei), as done in 
``optical'' Glauber calculations~\cite{Bialas:1976ed,Bialas:1977pd}. It should be noted that this method neglects some 
potentially important correlations between participating nucleon positions as will be discussed further in \Sect{cumulants}. 

A related, but fundamentally different approach is to distribute, event-by-event in a stochastic manner,  individual 
nucleons according to the 
smooth matter distribution and to evaluate the collision properties of the colliding nuclei by averaging over multiple 
events using Monte Carlo methods~\cite{Ludlam:1986dy,Shor:1988vk}. The key ingredients in \MCG\ calculations are the 
following: 
\begin{enumerate}
\item The nucleon position centers in each nucleus are distributed according to a probability distribution given
by the smooth nuclear density function, $\rho$.
One can think of the smooth nuclear density as a quantum mechanical single-particle probability distribution for 
the nucleon positions and their actual values in an individual collision event as a ``measurement'' of their positions
in a given collision experiment.
\item The nucleons are assumed to travel in straight trajectories along the beam direction throughout the reaction,
\ie~their transverse positions are ``frozen'' during the short time when the two high-energy nuclei pass through each 
other. 
\item The nucleons interact with nucleons in the oncoming nucleus by means of the nucleon--nucleon inelastic cross 
section~($\signn$) appropriate for the beam energies under consideration~(measured in proton-proton collisions).
The nucleon--nucleon collisions occur and produce particles independently, 
\ie~dynamical correlations among the 
nucleon positions in the multi-particle nuclear wave function are assumed to be negligible. The only correlations
in the model are of geometrical nature and due to clustering effects from the interaction process itself, as explained 
in \Sect{cumulants}.
\end{enumerate}
Commonly, a nucleon--nucleon collision in the reaction is defined to occur if the Euclidean transverse distance between 
the centers of any two nucleons is less than the ``ball diameter'',
\begin{equation}
\label{eq:balldiameter}
D=\sqrt{\signn/\pi}\,.
\end{equation}

More specifically, the steps of the PHOBOS Monte Carlo~\cite{Back:2004je} calculation for a single 
event are the following:
\begin{itemize}
\item {\bf Impact parameter selection:}
The two nuclei are separated in the $x$-direction by an impact parameter, $b$, chosen 
randomly according to $dN/db \propto b$ up to some large maximum~(\mbox{$b_{\rm max}\simeq20\,$fm$>2R_{A}$}). 
Thus, nucleus $A$ is defined to be centered at $\{x,y\}=\{-b/2,0\}$ in
the transverse plane, while nucleus $B$ is centered at $\{x,y\}=\{+b/2,0\}$. In addition, both 
nuclei are centered at $z=0$, since the longitudinal coordinate of each nucleus does 
not matter for the subsequent steps~\footnote{Throughout the paper, we will keep the common 
choice that the reaction plane, defined by the impact parameter and the beam direction, is given 
by the $x$- and $z$-axes, while the transverse plane is given by the $x$- and $y$-axes.}.

\item {\bf Makeup of nuclei:} 
For each nucleus, we loop over the number of nucleons, 
$N_A$ and $N_B$, and for each nucleon center point choose random, uniformly distributed
azimuthal and polar angles, as well as a radius sampled randomly according to the radial 
density distribution $\rho(r)$.
Additionally, to mimic excluded volume effects, one may require a minimum inter-nucleon 
separation distance~($d_{\rm min}$) between the nucleon centers of all nucleons in the nucleus.
This introduces a geometrical correlation among the nucleon positions.
In the construction of the nuclei, we make sure that the center-of-mass of the 
nuclei is correctly positioned, \ie~we achieve $\sum x_i=\pm b/2$, $\sum y_i=0$ and $\sum z_i=0$
by shifting all nucleon centers by the average offset determined after the positions of all nucleons
in the nucleus have been generated. Thereby we ensure that the nuclear reaction and the 
otherwise arbitrary MC frames coincide.

\item {\bf Collision determination:} 
For each nucleon in nucleus $A$, we loop over the nucleons in nucleus $B$.
If the 2-dimensional Euclidean distance $\sqrt{\Delta x^2+\Delta y^2}$ between 
the nucleon from $A$ and the nucleon from $B$ is less than $D$ as defined in 
\Eq{eq:balldiameter}, the number of collisions suffered by both nucleons is incremented 
by one. 
If no such nucleon--nucleon collision is registered for any pair of nucleons,
then no nucleus--nucleus collision occurred. Counters for determination of the 
total~(geometric) cross section are updated accordingly.
\end{itemize}

Having access to the number of collisions suffered by each nucleon according to this purely geometrical~(classical) prescription 
allows straightforward calculation of $\Npart$, the number of nucleons which are struck at least once, and $\Ncoll$, the total number 
of nucleon--nucleon collisions. The latter is defined as the sum of collisions suffered by nucleons in one nucleus with nucleons from 
the other one~(to avoid double counting). For every collision, the calculation keeps track of the position and status of each 
nucleon in the event, for later usage in the calculation of the spatial eccentricity or any other interesting quantity.

The default parameters of the PHOBOS Glauber calculations for \AuAu\ and \CuCu\ collisions at $\snn=200\,$GeV are listed in 
\Tab{tab:tab1}. The nucleon--nucleon inelastic cross section of $\signn=42\,$mb 
is from \Ref{Yao:2006px}, while the parameters for the Fermi distribution, 
\beqann
\rho(r) \propto \left (1+\exp\left( \frac{r-R}{a} \right) \right)^{-1}\,,
\eeqann
\ie~the nuclear radius $R$ and the skin depth $a$, are from \Ref{DeJager:1987qc}. 
The minimum inter-nucleon separation distance is set to $d_{\rm min}=0\,$fm,
\ie~we generally ignore geometrical correlations in the multi-nucleon wave function due to a hard core, since their
effect, especially on the participant eccentricity, is found to be small, as we will report in \Sect{density}.

\begin{table}[tb!f]
\caption{\label{tab:tab1}Default parameters used in the PHOBOS \MCG\ calculations for \AuAu\ and \CuCu\
at $\snn=200\,$GeV.}
\begin{ruledtabular}
\begin{tabular}{lccccc}
System & $\signn\,$[mb] & $N_{A/B}$ & $R\,$[fm] & $a\,$[fm] & $d_{\rm min}\,$[fm] \\
\AuAu\ & $42$          & $197$     & $6.38$   & $0.535$  & $0$ \\ 
\CuCu\ & $42$          & $63$      & $4.20$   & $0.596$  & $0$ \\ 
\end{tabular}
\end{ruledtabular}
\end{table}

\begin{table}[tb!f]
\caption{\label{tab:tab2}Baseline, minimum and maximum values of parameters in the PHOBOS \MCG\ calculations for 
\AuAu\ and \CuCu\ collisions used in sensitivity studies of the eccentricity definitions.}
\begin{ruledtabular}
\begin{tabular}{llllllll}
                              &      & \multicolumn{3}{c}{\AuAu}    & \multicolumn{3}{c}{\CuCu}    \\
\multicolumn{2}{l}{Parameter}        & Base     & Min     & Max     & Base     & Min     & Max     \\
$\signn$                      & [mb] &  $42$    & $30$    & $45$    & $42$     & $30$    & $45$    \\
$R$                           & [fm] &  $6.38$  & $6.25$  & $6.51$  & $4.22$   & $4.14$  & $4.30$  \\   
$a$                           & [fm] &  $0.535$ & $0.482$ & $0.586$ & $0.596$  & $0.536$ & $0.656$ \\ 
$d_{\rm min}$                 & [fm] &  $0.4$   & $0$     & $0.8$   & $0.4$    & $0$     & $0.8$   \\
\end{tabular}
\end{ruledtabular}
\end{table}

\begin{figure*}[t]
\begin{minipage}[t]{0.49\textwidth}
\includegraphics[width=\textwidth]{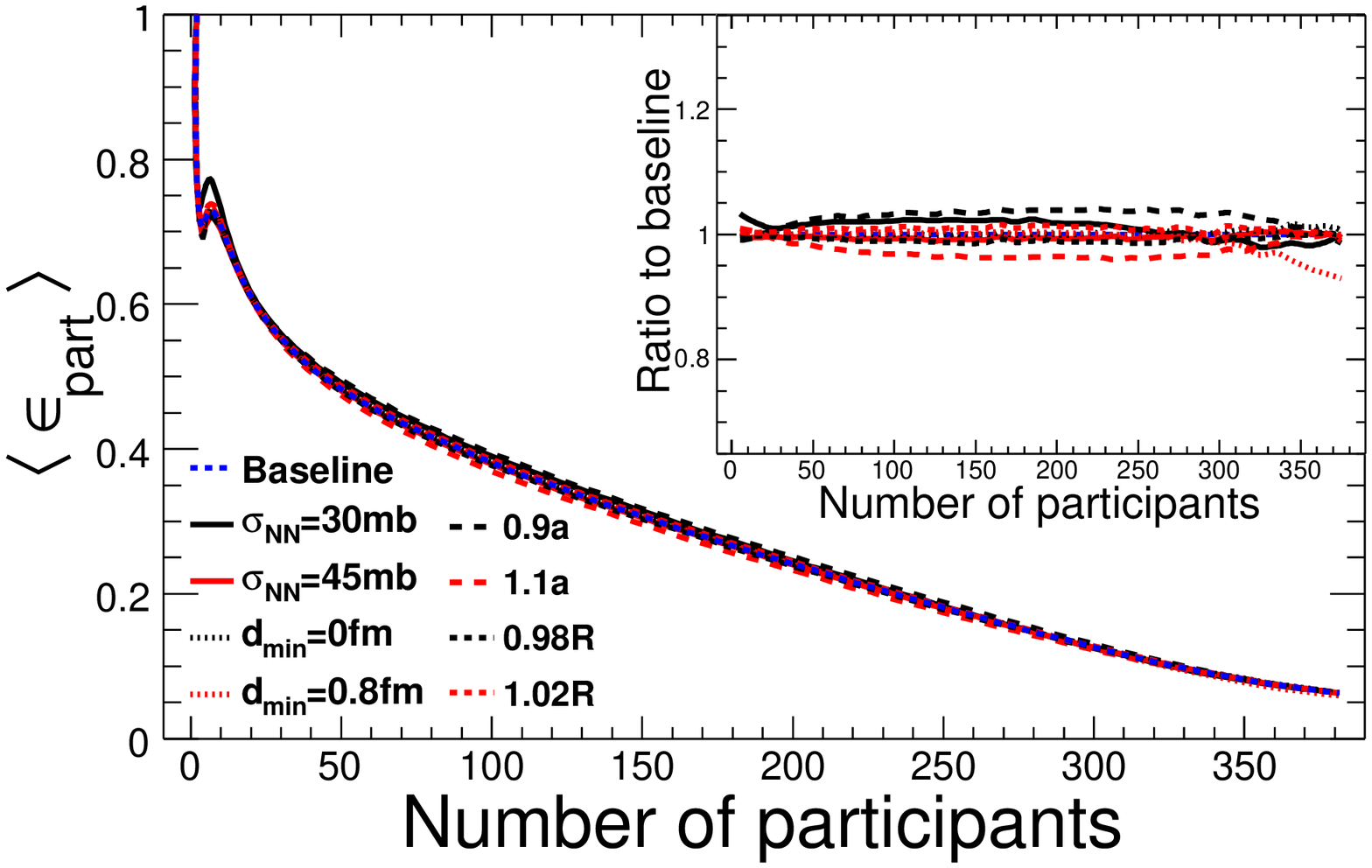}
\end{minipage}
\hspace{0.1cm}\hfill
\begin{minipage}[t]{0.49\textwidth}
\includegraphics[width=\textwidth]{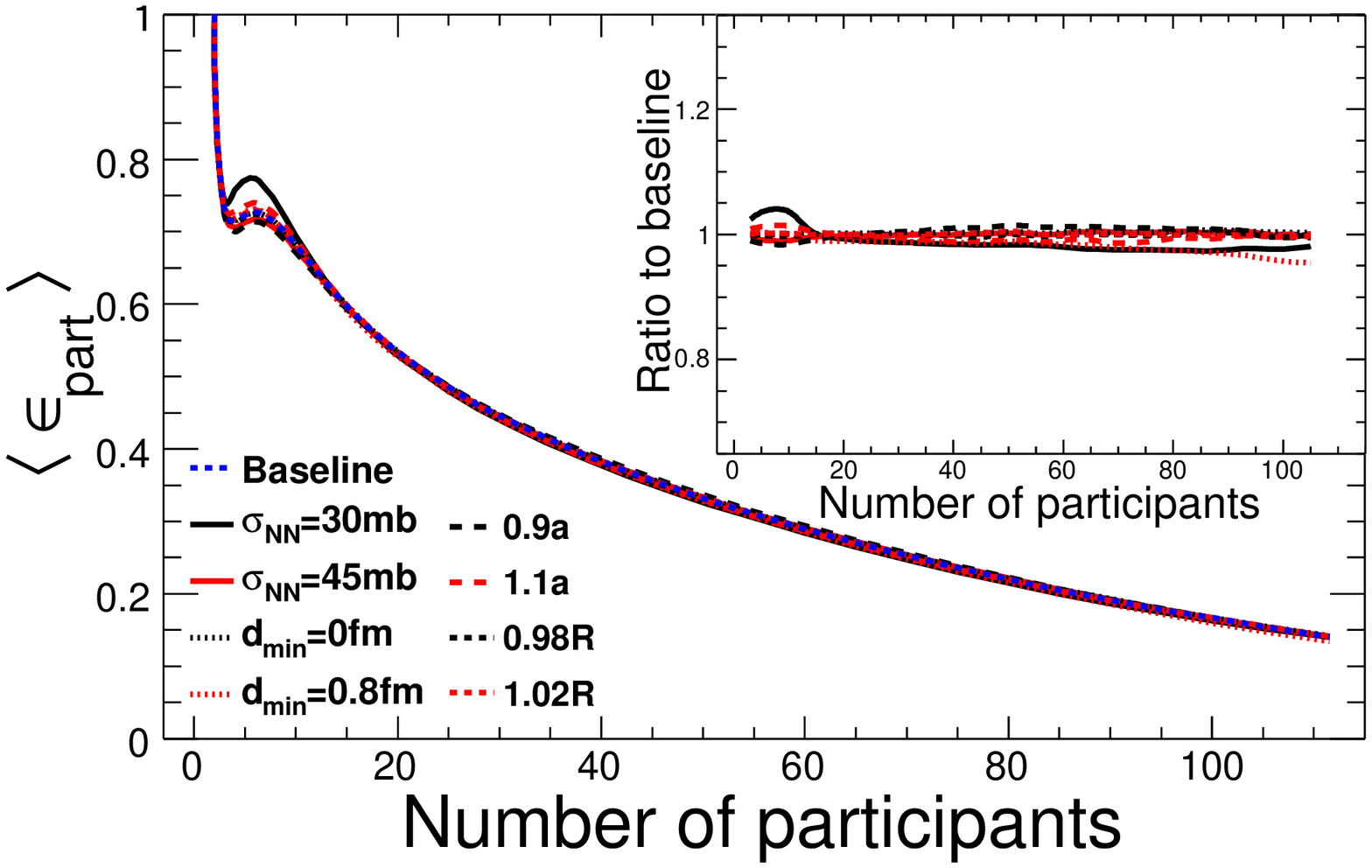}
\end{minipage}
\begin{minipage}[t]{0.49\textwidth}
\includegraphics[width=\textwidth]{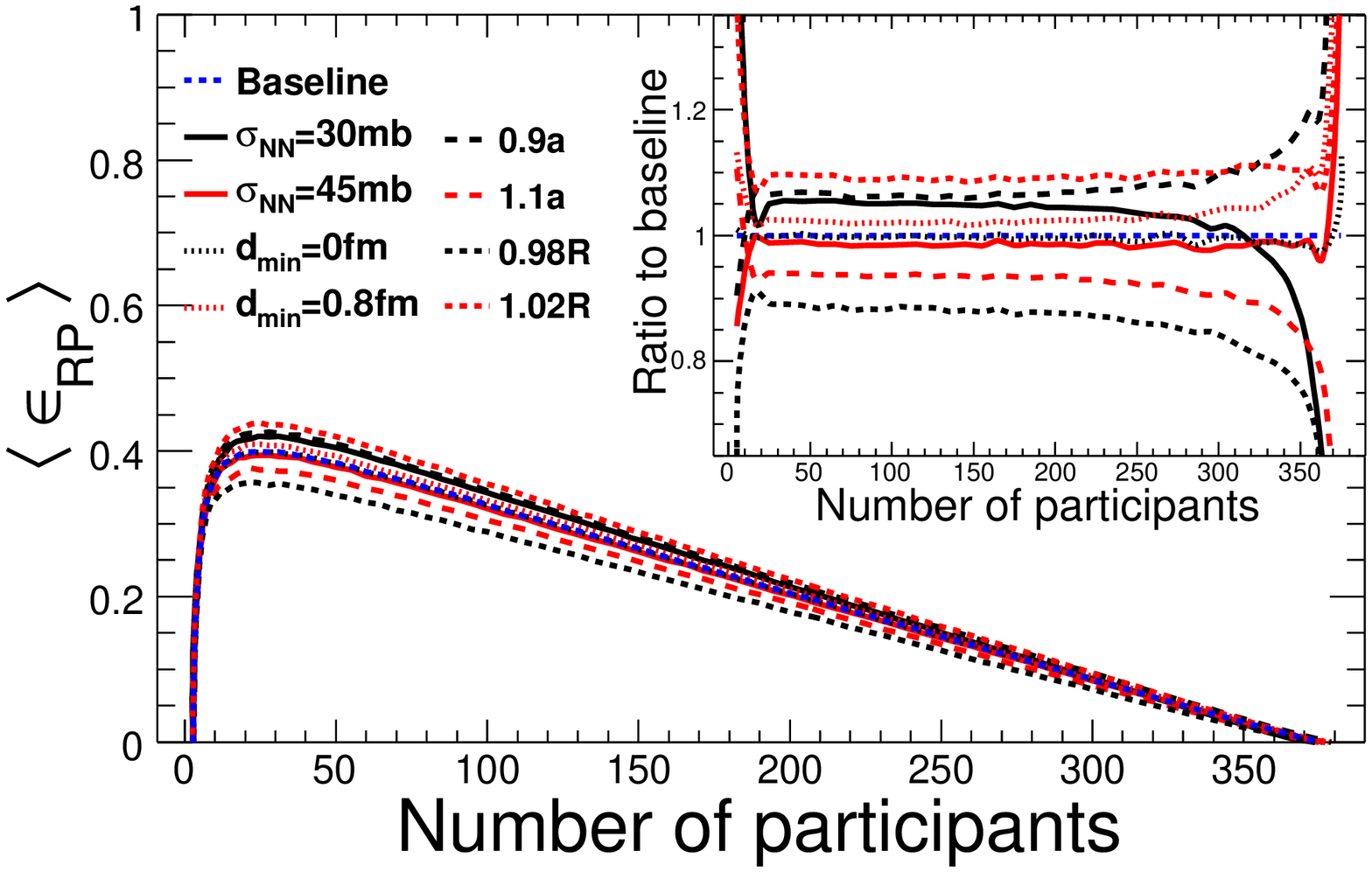}
\end{minipage}
\hspace{0.1cm}\hfill
\begin{minipage}[t]{0.49\textwidth}
\includegraphics[width=\textwidth]{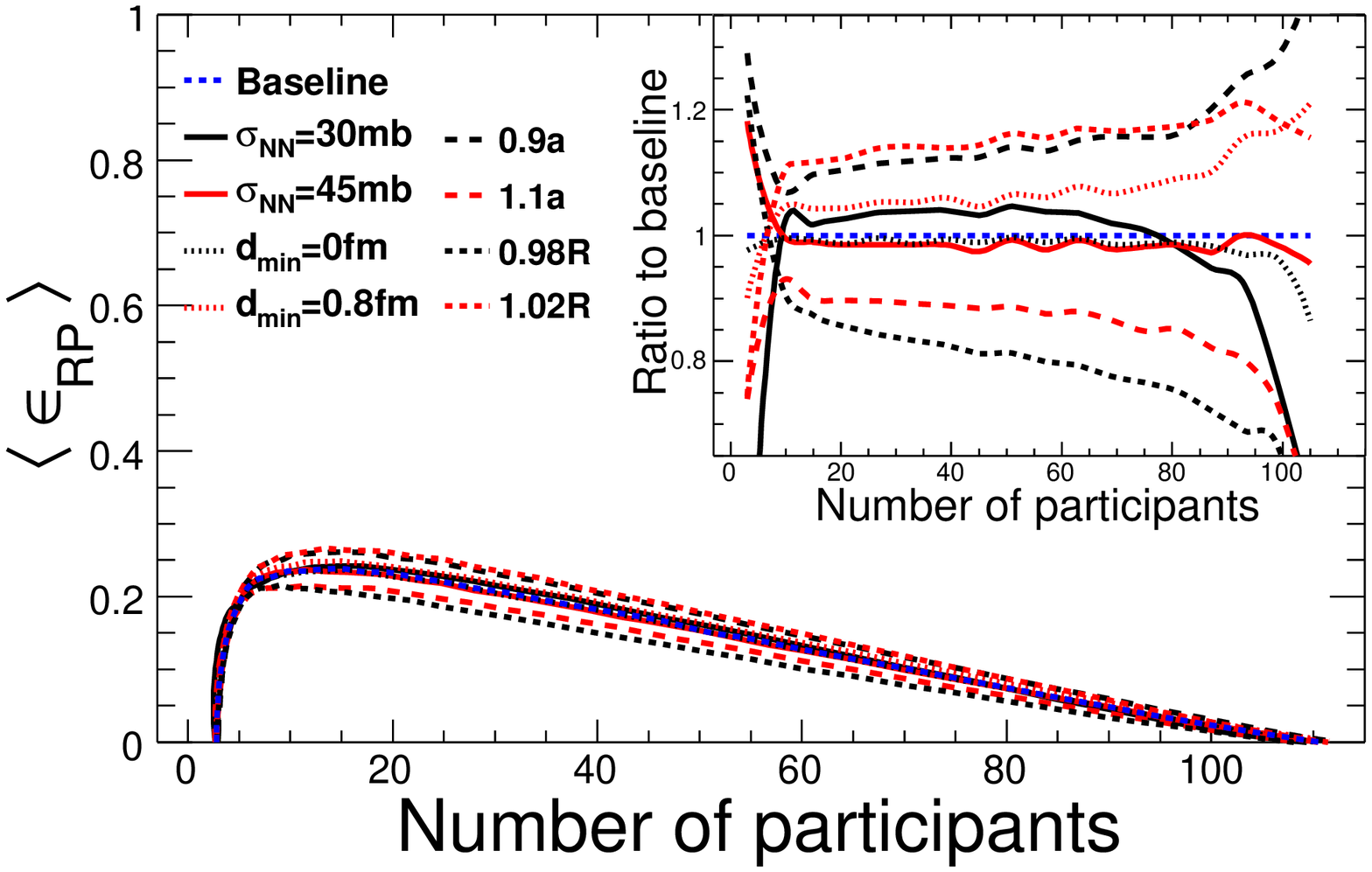}
\end{minipage}
\caption{\label{fig:eccrob}Participant~(top) and reaction plane~(bottom) eccentricity in \AuAu~(left) and
\CuCu~(right) collisions as a function of $\Npart$ for variations of the Glauber parameters as given
in \Tab{tab:tab2}. The inset shows the ratio \wrt\ the baseline calculation in each case.}
\end{figure*}

A few comments are in order explaining what this model delivers and how we subsequently interpret its output.
As described so far, the \MCG\ model records only the (transverse) position and collision status of each nucleon.
No particles are produced in the calculation and dynamical correlations among the nucleons in the nuclear wave 
function are neglected. By specifying the nuclear positions exactly, \ie~as long as we do not allow for a smearing
around the points given by the model, we are prohibited~(by quantum mechanical uncertainty) from imposing any constraints
on the momenta of the scattered nucleons and, in consequence, of the particles produced by the collision. Source 
eccentricities calculated directly from the distribution of~(exact) nucleon positions obtained from the \MCG\ model
can therefore, at least in principle, not immediately be assumed to represent the eccentricity of the produced matter
distribution which drives the anisotropy of the subsequent collective expansion. For this reason, we extend in 
\Sect{partprodmodels} the model by smearing the resulting nucleon--nucleon collision points with a profile function in
order to model the production of~(approximately thermalized) matter with finite temperature and restricted
particle momenta in the neighborhood of the collision points delivered by the \MCG\ model. The~(in-)sensitivity
of the source eccentricity of these matter distributions to the parameters of the smearing profile
is studied in detail.

\section{\label{robustness}Robustness of the eccentricity} 

In this section, we evaluate the effects of variations in the
nuclear density distributions and of various assumptions
about the sources and spatial localization of the initial matter 
distributions on the eccentricity and its centrality dependence.
The two definitions of eccentricity considered in this section are 
the reaction plane eccentricity~(see \Eq{eq:epsrp})
\[
\erp = \frac{\sigma^2_y-\sigma^2_x}{\sigma^2_y+\sigma^2_x}
\]
and the participant eccentricity~(see \Eq{eq:epspart})
\[
\epart = \frac{ \sqrt{ (\sigma^2_y-\sigma^2_x)^2 + 
4 \sigma_{xy}^2}} {\sigma^2_y+\sigma^2_x}
\] 
where $\sigma^2_x$, $\sigma^2_y$ and $\sigma_{xy}$ are the (co-)variances
of the participant-weighted nucleon distribution in a given \MCG\ event~\footnote{Both 
definitions have already been used in \Refs{Manly:2005zy,Alver:2006wh}.}.
Their definitions and relation to the standard eccentricity~($\es$,~\Eq{eq:epsstd})
of the event-averaged distribution, are discussed in \App{eccdef}.

\subsection{\label{density} Variation of Density Parameters}

Before \Refs{Miller:2003kd,Manly:2005zy} the purpose of \MCG\ 
calculations was to estimate global properties of nucleus--nucleus collisions, 
\ie~to calculate centrality- and eccentricity-related quantities {\it on average}, 
based on~(large) samples of Glauber events. Since both the participant eccentricity
and the reaction plane eccentricity, explicitly involve an interpretation of each 
\MCG\ event individually, it is important to understand their dependence on the choice 
of the \MCG\ calculation parameters.
A number of sources of systematic error are studied by varying a specific parameter \wrt\ the baseline parameter set as listed 
in \Tab{tab:tab2}. 
The baseline values for the sensitivity study correspond to the default parameter set for $\snn=200\,$GeV 
except for the minimum inter-nucleon separation distance which is here set to $d_{\rm min}=0.4\,$fm to 
match the default value in HIJING~\cite{Wang:1991ht}. We study the variation of all the main Glauber parameters except for the 
atomic mass number. The nuclear radius~($R$) is varied by $\pm2$\%, the nuclear skin depth~($a$) by $\pm10$\%; both variations 
are several times larger than the systematic error assigned to their measurements by the authors of \Ref{DeJager:1987qc}.
The nucleon--nucleon inelastic cross-section~($\signn$) is varied by more than the experimentally 
spanned region at RHIC~\footnote{A posteriori, this is justified since the dependence on $\signn$ turns out to be small. 
Thus, this approach will allow the treatment of the systematics at all collision energies in the same way.}.
The minimum inter-nucleon separation distance, which is not known experimentally, is varied by $\pm100$\%. 
As one can see in \Fig{fig:eccrob}, both eccentricity definitions are quite stable within the studied range of
Glauber parameters. In particular, this is true for the participant eccentricity. Not unexpectedly, the difference between the 
two definitions, $\epart$ and $\erp$, is most pronounced for \CuCu. We have also found that varying two parameters at the same 
time does not increase significantly the observed variation in eccentricity.

\subsection{\label{partprodmodels}Particle Production Models}

\begin{figure}[t]
\includegraphics[width=0.49\textwidth]{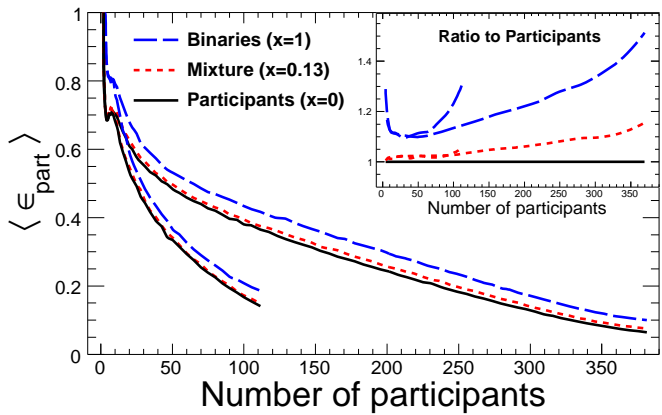}
\caption{\label{fig:bincoll}Comparison of participant versus binary weighted participant eccentricity, as well as the $x=0.13$
mixture, as a function of $\Npart$ in \AuAu~(upper set of curves) and \CuCu~(lower set of curves) collisions at $\snn=200\,$GeV. 
The inset shows the ratio \wrt\ the participant-weighted eccentricity.}
\end{figure}

\subsubsection{Binary Collisions versus Participants}
The observed particle multiplicity at mid-rapidity scales somewhat more strongly than linearly with the number of participating 
nucleons~\cite{Back:2004dy}. This can be parametrized by postulating a second~(smaller) contribution to particle 
production that scales with the number of binary nucleon--nucleon collisions~\cite{Kharzeev:2000ph}. The \MCG\ 
model as described above, can be extended to implement matter distributions produced according to a ``two-component''
scenario, where some of the matter is generated proportionally to the number of binary collisions. 
To add this feature, it is necessary to define two origins of matter production:
\begin{itemize}
\item Participant nucleons, which create the matter by means of ``exciting'' the nucleon. 
\item Binary nucleon--nucleon collisions, which create the matter locally via a two-body interaction.
\end{itemize}

The latter mechanism suggests that the produced matter, to be incorporated into the calculation of spatial eccentricity, should be
centered between the colliding nucleons. We achieve this by means of a ``pseudo-particle'' that is located at the center of mass
of the pair of colliding nucleons and keeps track of the ``collision-weighted'' matter with an appropriate weight~$x$. 
Thus, contributions to the eccentricity from 
participants are weighted by $\frac{1-x}{2}$ while those from binary collisions by $x$.
In \Fig{fig:bincoll}, the results for purely participant~($x=0$) and purely collision~($x=1$) weighted participant eccentricity are 
shown and compared to the case $x=0.13$ which has been found to describe the centrality dependence of the multiplicity at mid-rapidity
according to $\dncde = \dnppcde\,\left[\frac{1-x}{2} \, \Npart + x \, \Ncoll\right]\,$\cite{Back:2004dy}. 
Independent of centrality, the collision weighted $\epart$ values are shifted to larger eccentricity
for all centralities, similar to what is known for the standard eccentricity~\cite{Kolb:2001qz}.
The presumably realistic case of $x=0.13$ yields about $10$\% larger eccentricity in the most central \AuAu\ 
collisions.

\begin{figure}[t]
\includegraphics[width=0.49\textwidth]{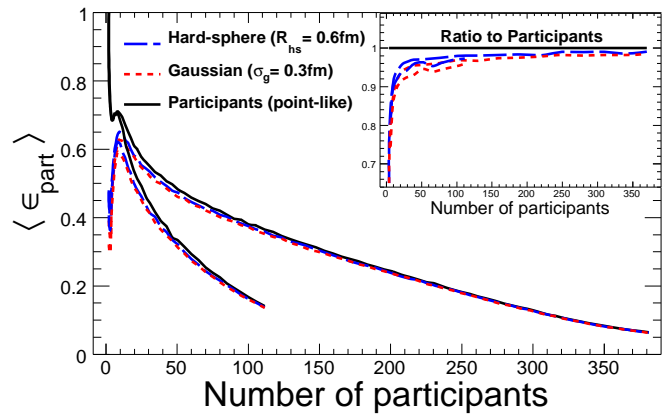}
\caption{\label{fig:smeared}Comparison of participant eccentricity as a function of $\Npart$ in \AuAu~(upper set of curves) and 
\CuCu~(lower set of curves) collisions at$\snn=200\,$GeV for point-like, hard-sphere, and Gaussian matter distributions. 
The inset shows the ratio \wrt\ the point-like participant eccentricity.}
\end{figure}

\subsubsection{Effects of Smeared Matter Distributions}
The driving force for hydrodynamic elliptic flow is not directly the eccentricity of the 
distribution of participating nucleons or binary nucleon--nucleon collisions, but rather 
the anisotropy of the pressure gradients in the initially produced hot matter after 
it has thermalized. The transverse distribution of this matter will be smeared around the 
transverse positions of the participating nucleons or binary collision points. 
To study the behavior of the eccentricity under different definitions,
we introduce a general procedure for incorporating a variety of matter density distributions.
In this approach, the contribution of each matter production point~(\ie~the center of a participant nucleon or a binary collision) 
at~($x_i,y_i,z_i$) is smeared according to $P(x-x_i,y-y_i,z-z_i)$, leading to a continuous weight function defined at all space-points 
in the transverse plane, $w(x,y,z)=\sum\,P(x-x_i,y-y_i,z-z_i)$. Averages and higher moments in space-time are then calculated~(for 
individual events) using this weight function, \eg~$\{x\}=\int x\, w(x,y,z) \, \dd x \dd y \dd z$. The point-like \MCG\ cases described 
above correspond to the choice of \mbox{$P(x-x_i,y-y_i,z-z_i)=\delta(x-x_i)\,\delta(y-y_i)\,\delta(z-z_i)$}. We look at two different, 
azimuthally symmetric, parametrizations for the smearing profile
\begin{itemize}
\item Hard-sphere smearing, $P_{\rm hs}(r)\propto\,r^2\,\theta(R_{\rm hs}^2-r^2)$
\item Gaussian smearing, $P_{\rm g}(r)\propto\,r^2\,\exp\left(-r^2/2\sigma_{\rm g}^2\right)$
\end{itemize}
where $r^2=(x-x_i)^2+(y-y_i)^2+(z-z_i)^2$ and $\theta$~denotes the step function.
In the following, we estimate meaningful choices for the parameters, $R_{\rm hs}$ and 
$\sigma_{\rm g}$. For $R_{\rm hs}$, it makes sense to use the interaction radius, $R_{\rm hs}=D/2=\sqrt{\signn/4\pi}$, 
since the nucleons in the Glauber model are assumed to interact if their centers are within the ``ball diameter''~$D$, 
\Eq{eq:balldiameter}. For $200\,$GeV, this corresponds to $R_{\rm hs}\approx0.6\,$fm. Then, matching the root-mean-square~(RMS) width of the 
Gaussian distribution to the RMS width of the hard-sphere, $\sigma_{\rm g}^2=R_{\rm hs}^2/5$, leads to $\sigma_{\rm g}\approx0.3\,$fm. 
In \Fig{fig:smeared} we show the results for the participant 
eccentricity calculated for point-like, hard-sphere, and Gaussian local matter distributions, using this set of 
parameters. The comparison reveals that for both collision systems the way the produced matter is distributed around the \MCG\
interaction points does not significantly influence the observed value of the eccentricity except for extremely small systems. 
This also shows that the quantum mechanical uncertainty on the transverse positions of the interaction points has no major 
influence on the initial source eccentricity.
Note that similarly to what was reported in \Ref{Broniowski:2007ft} we find significantly different centrality dependence 
for $\epart$ if we allow smearing out the local matter sources to a very large extent. For example, $\epart$ in \AuAu\ collisions 
at all centralities does not exceed $0.15$ for Gaussian smearing with $\sigma_{\rm g}=2\,$fm. 

\section{Correlations and Fluctuations} 

In this section, we focus on eccentricity cumulants~(which enter the discussion
and interpretation of elliptic flow data) in the context of fluctuating initial 
conditions and for different realizations of the Glauber model initial state.

\subsection{\label{sensitivity}Sensitivity of the Event-Plane Method\\
to Underlying Flow Fluctuations}

There are different ways of extracting the elliptic flow from data: the event-plane 
method, two-particle correlations, multi-particle cumulants, \etc\ (see 
\Refs{Poskanzer:1998yz,Borghini:2001vi}). Each flow measurement is based on a 
different moment of the final-particle momentum distribution and thus is differently
affected by event-by-event flow fluctuations (and non-flow correlations).
If elliptic flow is proportional to the spatial anisotropy, the eccentricity 
scaling should be performed with corresponding moments of the participant 
eccentricity~\cite{Miller:2003kd,Bhalerao:2006tp}.

It has been explicitly stated~\cite{Bhalerao:2006tp}~(see also~\cite{Borghini:2001vi}) 
that the event-plane method~($\vrp$), used by the PHOBOS experiment to measure elliptic 
flow, really measures $\sqrt{\av{v_2^2}}$ rather than $\av{v_2}$, \ie~the RMS 
rather than the mean of $v_2$. More specifically, it has been claimed that 
$\av{v_2} \le \vrp \le \sqrt{\av{v_2^2}}$ depending on the event-plane resolution, with 
the upper limit being approximately reached under RHIC conditions~\cite{Ollitrault:2006privat}.
Here, $\av{\dots}$ indicates an average over many collision events. 
In the following, we will investigate and confirm this claim.

In the event-plane method~\cite{Poskanzer:1998yz}, one uses the particles from one side of the 
detector~(subevent $A$) to estimate the event plane, the plane relative to which the flow develops, 
given by 
$\Psi_2^{A}= \frac{1}{2} \tan^{-1} \left[\sum\sin(2\phi_i)/\sum\cos(2\phi_i)\right]$.
One then correlates the particles from the other, symmetric, side of the detector~(subevent $B$) 
with this event plane to obtain the uncorrected flow signal for the given subevent, 
$v^{\rm obs}_{2,B}=\{\cos\left(2\phi_i-2\Psi_2^A\right)\}$. 
Here, as before and in the appendices, $\{\dots\}$ indicates the average over an individual event.
The roles of subevents $A$ and $B$ can then be interchanged to obtain $\Psi_2^B$, $v^{\rm obs}_{2,A}$,
and thus the observed flow signal for the whole event $v_2^{\rm obs}$. 
Assuming small dynamical and non-flow correlations one has (for symmetric subevents)
\beq
 \av{v_{2,B(A)}^{\rm obs}} = \av{v_2} \, \av{\cos\left(2\Psi_2-2\Psi_2^{A(B)}\right)}
\eeq
where the average is, unlike before, not over particles in a given event but over events in the given 
centrality, $\eta$ and $\pt$ bin. In this equation, $\Psi_2$ stands for the actual event plane 
angle, which defines the orientation of the $v_2$ signal in a particular event~\cite{Alver:2006wh},
and $\av{\cos\left(2\Psi_2-2\Psi_2^{A(B)}\right)}\equiv R$ quantifies the 
event-plane resolution, which itself depends on $v_2\,$\cite{Poskanzer:1998yz}.
The resolution can be estimated based on data alone, 
\beq
 R=\sqrt{\av{\cos\left(2\Psi_A-2\Psi_B\right)}}
\eeq
leading to \mbox{$\av{v_{2}^{\rm obs}} = \av{v_2} R$}.
The presence of a fraction~($f_{\rm bkg}$) of uncorrelated background particles in addition to the $N$ particles 
that carry the flow signal corresponds~(restricting ourselves to second harmonic contributions) to a distribution of
\beq
 \frac{\dd N}{\dd\phi} =(1 + \fbkg) \frac{N}{2\pi} \, \left[1+\frac{2v_2}{1 + \fbkg} \cos\left(2\phi-2\Psi_2\right)\right ]\,,
\eeq
leading to an apparent suppression of the observed flow signal by $1+\fbkg$. Correcting for this effect, we arrive at the 
final expression for the flow measured via the event-plane method
\beq
 \label{eq:v2rp}
 \vrp \equiv \av{v_{2}} = \frac{1+\fbkg}{R} \,\av{v_{2}^{\rm obs}}\,.
\eeq
This includes the two main experimental corrections, namely for suppression and event-plane resolution~\cite{Back:2002gz}.

From \Eq{eq:v2rp} it is not obvious if the $\vrp$ scales with the mean or the RMS of an underlying $v_2$ distribution, and
how such a behavior depends on $R$. However, one can study this behavior numerically with a MC calculation that creates 
events with $v_2$ distributed according to a given distribution, $P(v_2)$. For every event, we take $N$ particles~(at 
mid-rapidity) to carry the flow signal, according to $\dd N/\dd\phi\propto 1+2v_2\cos(2\phi)$, and $\fbkg\,N$ particles~($\fbkg\ge0$) 
to represent the uncorrelated background, which are added with a uniform azimuthal distribution. 
To relate the obtained $\vrp$ to a moment of the input distribution, we implicitly define 
the exponent $\alpha$ according to 
\beq
 \label{eq:alpha}
 \left\langle\!\left\langle v_2^{\alpha} \right\rangle\!\right\rangle^{\frac{1}{\alpha}} = \vrp 
\eeq
where the ensemble average $\left\langle\!\left\langle \cdots \right\rangle\!\right\rangle$ is calculated with the
underlying $v_2$ distribution, $P(v_2)$. 
One obtains the sensitivity of the event-plane method to $P(v_2)$: \mbox{$\alpha=1$} corresponds to 
a scaling of $\vrp$ with the mean, \mbox{$\alpha=2$} to a scaling with the RMS of $P(v_2)$, also often 
denoted $v_2\{2\}\,$\cite{Borghini:2001vi}. 
In our calculation, we choose $v_2$ to be uniformly distributed between $v_{2, {\rm min}}$ and 
$v_{2,{\rm max}}$, \ie~the true mean and RMS are $0.5\,(v_{2,{\rm min}}+v_{2,{\rm max}})$ and 
$\sqrt{(v^2_{2,{\rm min}}+v_{2,{\rm min}}v_{2,{\rm max}}+v^2_{2,{\rm max}})/3}$, respectively. 

\begin{figure}[tb]
\includegraphics[width=0.49\textwidth]{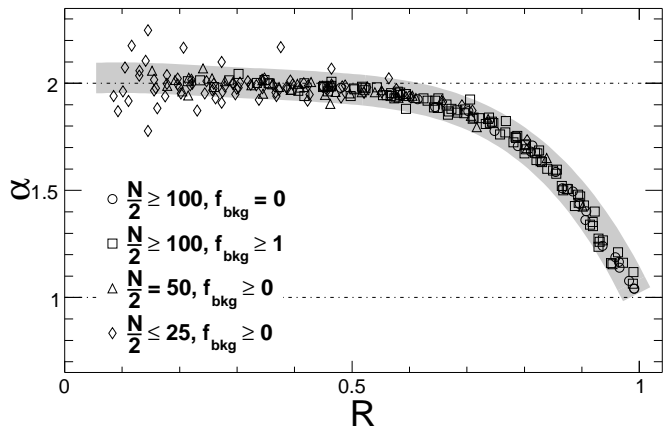}
\caption{\label{fig:alphares}Dependence of $\alpha$, \Eq{eq:alpha}, on the event-plane resolution, $R$, for $P(v_2)$ uniform
with various combinations of $v_{2,{\rm min}}$ and $v_{2,{\rm max}}$, number of signal~($N$) and background~($\fbkg N$) particles 
in the MC calculation~(see text for more details). 
The shaded band covers the parameter errors obtained from a polynomial fit to the data.}
\end{figure}

\Fig{fig:alphares} shows the result of the calculation for various
combinations of $v_{2,{\rm min}}$ and $v_{2,{\rm max}}$ between $0.01$ and $0.3$, various values of $N$ between $30$ and $1000$, 
as well as values of $\fbkg$ between $0$ and $5$. For each set of parameters, $10^{7}$ events have been simulated. 
The number of particles in each event is not allowed to fluctuate, \ie~exactly $N+\fbkg N$ particles are created in every event. 
Within a reasonable spread that increases with decreasing resolution, the $\alpha$ values are found to lie on a common curve as 
a function of $R$, with no or weak dependence at most on the chosen simulation parameters. 
The PHOBOS $\vrp$ measurements lie in the range of $0.15\lsim R \lsim 0.55$ for \AuAu\ and 
$0.13 \lsim R \lsim 0.33$ for \CuCu\, where $\vrp$ scales approximately with the RMS of the underlying $v_2$ distribution.
This result is supported by recent, full detector simulations with HIJING MC events that incorporate known, but fluctuating, 
$v_2$ values. For the measured value of $\sigma_{v_2}/\avsm{v_2}\approx40$\%~\cite{Alver:2007qw} this implies that
the PHOBOS $\vrp$ measurements are about 10\% larger than the mean elliptic flow. 

\subsection{\label{parteccscaledv2}Participant-Eccentricity-Scaled Elliptic Flow}

It is expected~\cite{Heiselberg:1998es,Voloshin:1999gs} that $v_2/\myepsilon$ scales with the
transverse charged particle area density at mid-rapidity according to 
$v_2/\myepsilon \propto 1/S \, \dncdy$, where $S$ is the overlap area 
in the transverse plane. The data are measured in bins of centrality, 
translating into
\beq
 \left. 
 \frac{\vrp}{\epsrp}
 \propto \frac{1}{\av{S}} \, 
 \av{\dncdy}\, \right|_{\abs{y}\le1}\,.
\eeq
Here, $\vrp$ is the PHOBOS estimate of the ensemble-averaged elliptic flow 
according to \Eq{eq:v2rp}, and $\epsrp$ is a suitable ensemble-averaged initial 
source eccentricity. In view of the discussion of the preceding subsection,
$\vrp$ is about $v_2\{2\}$, the RMS of the underlying $v_2$ distribution. Following
the suggestion of \Refs{Miller:2003kd,Bhalerao:2006tp} we therefore scale it with
$\epsrp = \eptwo \equiv \sqrt{\avsm{\epart^2}}$, the RMS of the participant
eccentricity distribution obtained from the \MCG\ model. 
The overlap area
\[
S=\pi\sqrt{\sigma^2_x\sigma^2_y-\sigma^2_{xy}}
\]
corresponds to the area of the tilted overlap ellipse~(see \App{eccdef}, and 
especially \Eq{eq:overlaps}).

\begin{figure}[t]
\includegraphics[width=0.49\textwidth]{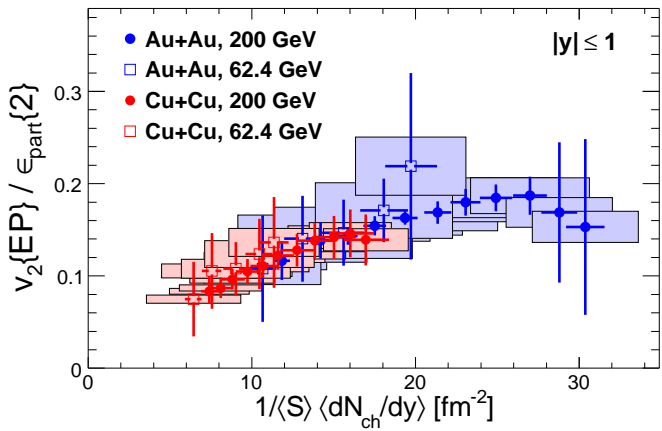}
\caption{\label{fig:scaledv2}Participant-eccentricity-scaled elliptic flow versus
transverse charged particle area density at mid-rapidity for \CuCu\ and \AuAu\ collisions 
at $\snn=62.4$ and $\snn=200\,$GeV. The horizontal and vertical error bars originate
from the combined statistical and systematic errors of the data~($90$\%~C.L.). The
shaded boxes are the result of the systematic errors assigned to $\eptwo$ and 
$S$ by the variation of the \MCG\ parameters~($90$\%~C.L.).}
\end{figure}

To arrive at the centrality average for $\eptwo$ 
and $S$, we fold their $\Npart$ dependence with the distribution of 
$\Npart$ values obtained for each centrality bin from full detector simulations.
Independent of species and energy, we scale the elliptic flow data by $0.9$ and 
the mid-rapidity yields by $1.15$ to convert from pseudo-rapidity~($\eta$) 
to rapidity~($y$) as described in \Ref{Adler:2002pu,Back:2004zg}. 
\Fig{fig:scaledv2} shows the result
for \CuCu\ and \AuAu\ collisions at $62.4$ and $200\,$GeV. The flow data are 
from \Refs{Back:2004mh,Alver:2006wh}, and the mid-rapidity yields from 
\Refs{Back:2002uc,Back:2005hs,alver:2007we}. 
We distinguish two types of errors that are individually propagated in the
error calculation of the ratios: 
a)~Systematic and statistical errors (if available) from data added in 
quadrature to obtain total $90$\%~C.L., and b)~systematic errors~($90$\%~C.L.) 
assigned to the MC quantities obtained by the variation of Glauber 
parameters \wrt\ the individual baseline values~(cp.~\Tab{tab:tab2}). 
As reported earlier~\cite{Manly:2005zy,Alver:2006wh}, we find a common
scaling between the different systems. However, within the errors, it is 
difficult to tell whether the almost linear rise of the eccentricity-scaled 
elliptic flow 
breaks down at larger values of the area density which might indicate 
that the hydrodynamic limit is being reached at the top RHIC energy.

\subsection{\label{cumulants}Cumulants and Correlations}

As mentioned above, it is suggested~\cite{Miller:2003kd} that higher order 
cumulant moments of $v_2$ should be proportional to analogously defined higher 
order cumulant moments of the eccentricity, including:
\beqa
\myepsilon\{2\}^2 & \equiv & \av{\myepsilon^2} \nonumber \\
\myepsilon\{4\}^4 & \equiv & 2\av{\myepsilon^2}^2 - \av{\myepsilon^4}.
\eeqa

In \Ref{Bhalerao:2006tp} Bhalerao and Ollitrault~(B\&O) attempted to
derive expressions for $\eptwo$ and $\epfour$ semi-analytically,
making use of two strong approximations.  
First, the paper contains
an implicit assumption that all of the participant positions are
independent samples of some underlying distribution, or at least that
any correlations between participants do not affect the eccentricity
fluctuations~\cite{Ollitrault:2006privat}.  
Second, the expressions
in \Ref{Bhalerao:2006tp} were obtained using a Taylor expansion,
leading to a power series in $1/\Npart$ which is then truncated at
$1/\Npart$. Based on these approximations, they concluded that $\epfour$ 
is numerically equal to the standard eccentricity $\es$, vanishing
for central~($b=0$) collisions. 
This would in turn imply that higher order cumulants of the flow such as 
$v_2\{4\}$ are insensitive to fluctuations in the participant distribution.  
In this section we show that B\&O's assumptions are too strong and that
$\epfour$ for \CuCu\ collisions differs significantly from $\es$ when better 
approximations are made, especially when the role of correlations is taken 
into account, \eg\ for the usual PHOBOS \MCG\ calculations.

\begin{figure}[t]
\includegraphics[bb=30 20 485 750, width=0.4\textwidth]{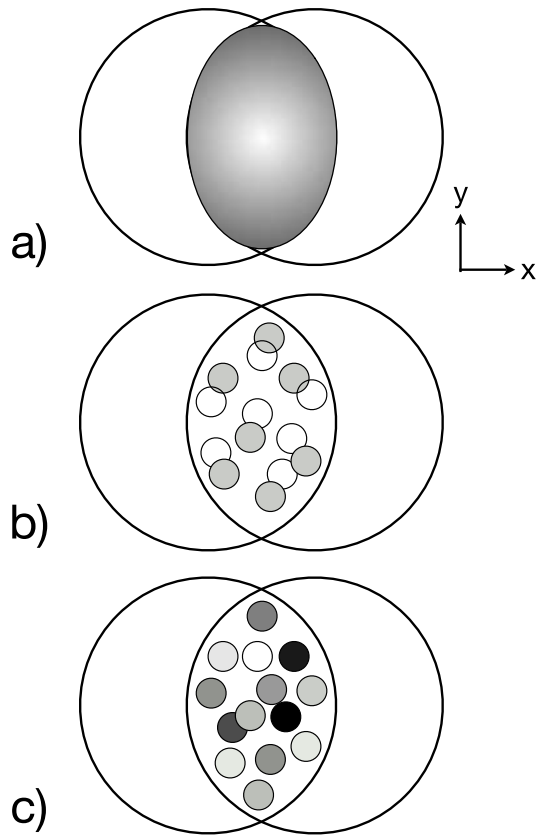}
\caption{\label{fig:glauber-diagram}
Schematic of densities for the different approaches: 
a)~optical limit, 
b)~full \MCG\ with correlated participants originating 
from each of the two nuclei in one \MCG\ event
and c)~mixed-event \MCG\ with uncorrelated participants
where every participant originates from an individual 
nucleon--nucleon collision obtained in a different
\MCG\ event.}
\end{figure}

\begin{figure*}[t]
\begin{minipage}[t]{0.49\textwidth}
\includegraphics[width=\textwidth]{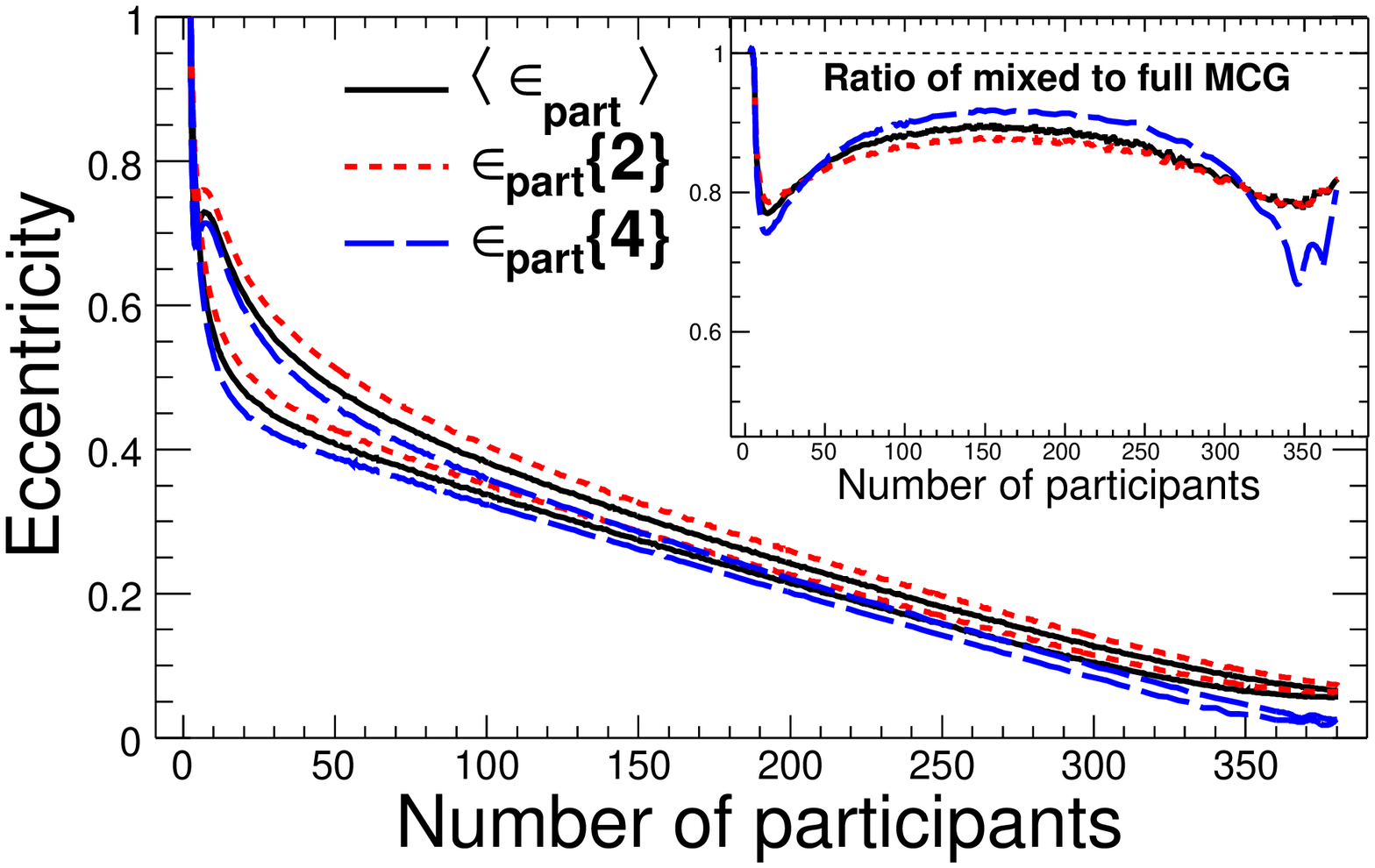}
\end{minipage}
\hspace{0.1cm}\hfill
\begin{minipage}[t]{0.49\textwidth}
\includegraphics[width=\textwidth]{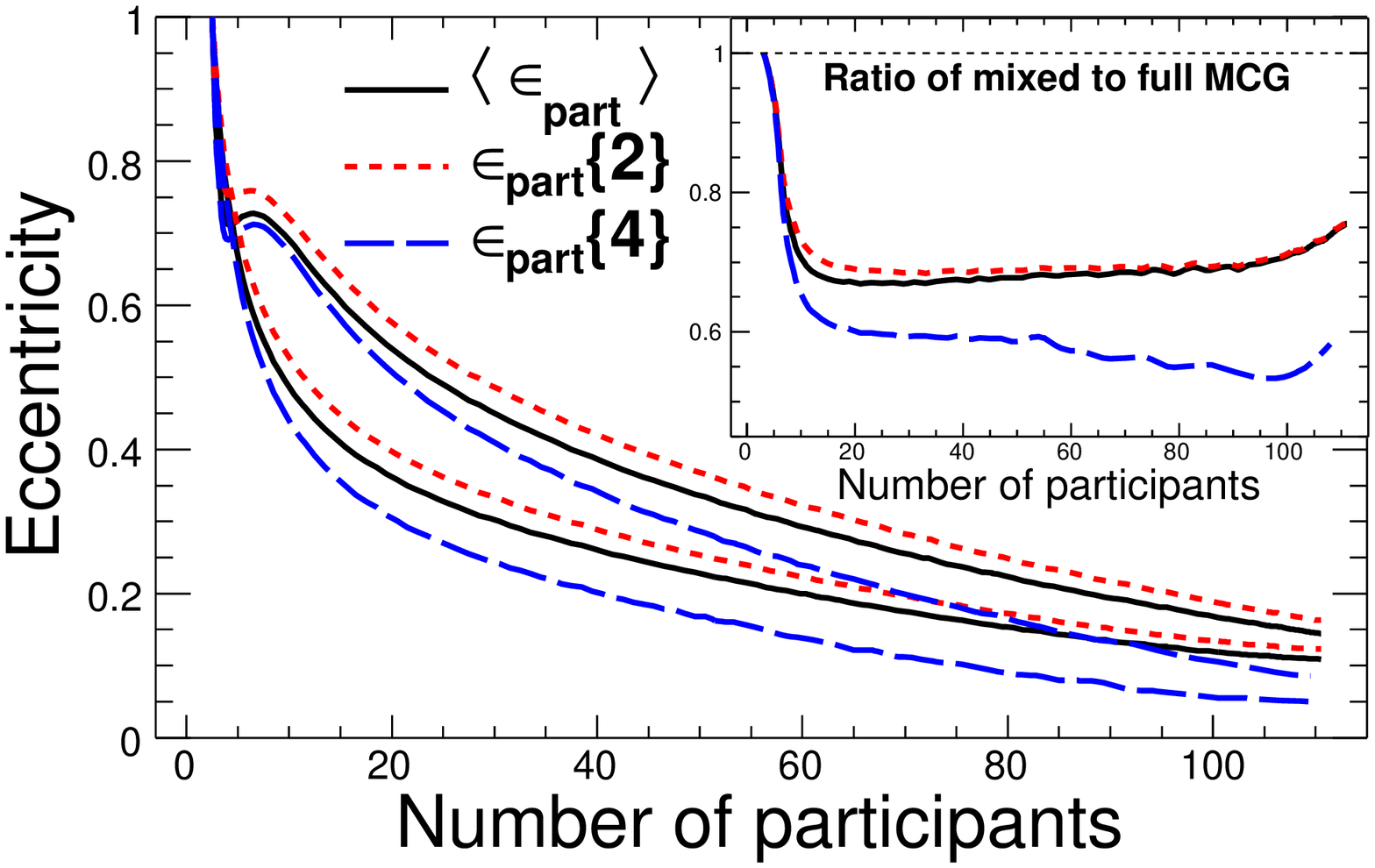}
\end{minipage}
\caption{\label{fig:eccrat}Participant eccentricity and cumulants, $\epart$, $\eptwo$ and $\epfour$, as a function of 
$\Npart$ in \AuAu~(left) and \CuCu~(right) collisions at $\snn=200\,$GeV for the full~(upper set of curves) and 
mixed-event~(lower set of curves) \MCG\ calculations. 
The inset shows the ratio of mixed-event to the full \MCG\ results.}
\end{figure*}

\subsubsection{Correlations of Nucleons in the Initial State}
The first approximation in \Ref{Bhalerao:2006tp} is to ignore the 
correlations between nucleon participant positions, even though we know 
that there are at least three sources for such correlations. 

First, in order to contribute to the produced matter, the participating 
nucleons must hit each other, which causes a correlation. For instance, 
in the case of a peripheral collision with two or three participants, the 
overlap region of the nuclei may be something like \mbox{$3\,$fm$\,\times 
1\,$fm} but the nucleons will necessarily all be within about $1\,$fm of
each other or else they would not be participants. In general, the
participant positions will tend to be more clustered in position space
than a random distribution since each participating nucleon must hit
another one. 

Second, if there are two nucleons from a given nucleus
which are close together in transverse position, then they will have a
tendency to be both hit or neither hit. Again this will contribute to
clustering of participant positions. 

The usual, full PHOBOS \MCG\ calculation takes both of 
these effects into account automatically, but the analytical expressions,
given in \Ref{Bhalerao:2006tp} and in \App{calccumulants} here, do not.  
The two different approximations are illustrated in 
\Fig{fig:glauber-diagram}~(panels~a) and~b)).

It should be noted that
a precise integral Glauber calculation at fixed impact parameter
should involve a complete $2\,(A+B)$-dimensional integral covering all
possible transverse positions of all nucleons involved in the
collision~\cite{Czyz:1969jg,Bialas:2001privat,Miller:2007ri}. Such a
formulation would include the same correlations that are covered
automatically by a \MCG. The usual practise of approximating the
Glauber integral with a single 2-dimensional integral~(optical Glauber 
model approximation) is just an approximation which neglects 
all the correlations described above. 

A third type of correlation would be genuine nucleon--nucleon position
correlations in the wave functions of the nuclei that, as mentioned
before, generally are ignored in all Glauber model calculations. 

\subsubsection{Uncorrelated Glauber Monte Carlo}
In order to estimate the role of pairwise spatial correlations among
the participant nucleons in a nucleus--nucleus collision,
we make use of a modified version of the \MCG\ code.  
In the modified version, at first a normal event is calculated of which
we record the impact parameter and corresponding number of participants.
Then, a mixed event is constructed from $\Npart$ independently calculated events, 
which are all required to have the same global characteristics, \ie~the same 
impact parameter and number of participants. 
From each such event, we choose one of the participating
nucleons, such that in the constructed mixed event none of the participating
nucleons is correlated to any of the other participants. The resulting 
approximation of the overlap density is illustrated in panel~c) of 
\Fig{fig:glauber-diagram}.

\Fig{fig:eccrat} shows the comparison of the full PHOBOS \MCG\ calculation
with the mixed-event \MCG\ calculation for the participant eccentricity
and its first two cumulants. We find that the contribution of correlations 
to $\epart$, $\eptwo$ and $\epfour$ is quite important for the smaller system, 
\ie~it is about $20$--$45$\% for \CuCu\ and about $5$--$10$\% for \AuAu\, 
and rather constant over a wide range of centrality. 
Note that the structure seen in \Fig{fig:eccrat}
at very low $\Npart$ values~(see also Figs.~\ref{fig:eccrob},\ref{fig:bincoll} 
and \ref{fig:smeared}) is genuine for the full \MCG\, and not present 
in the mixed-event case. 
Furthermore, spatial correlations among the participants in the initial state 
modeled by \MCG\ are found to be less important for the reaction plane and 
standard eccentricity definitions~(not shown). 
For both definitions the uncorrelated cases lead to slightly larger eccentricities.
The deviation between full and mixed-event \MCG\ decreases with increasing centrality
and is less than a few percent for the \AuAu\ and about $5$--$15$\% for the \CuCu\ system.

\begin{figure*}[t]
\begin{minipage}[t]{0.49\textwidth}
\includegraphics[width=\textwidth]{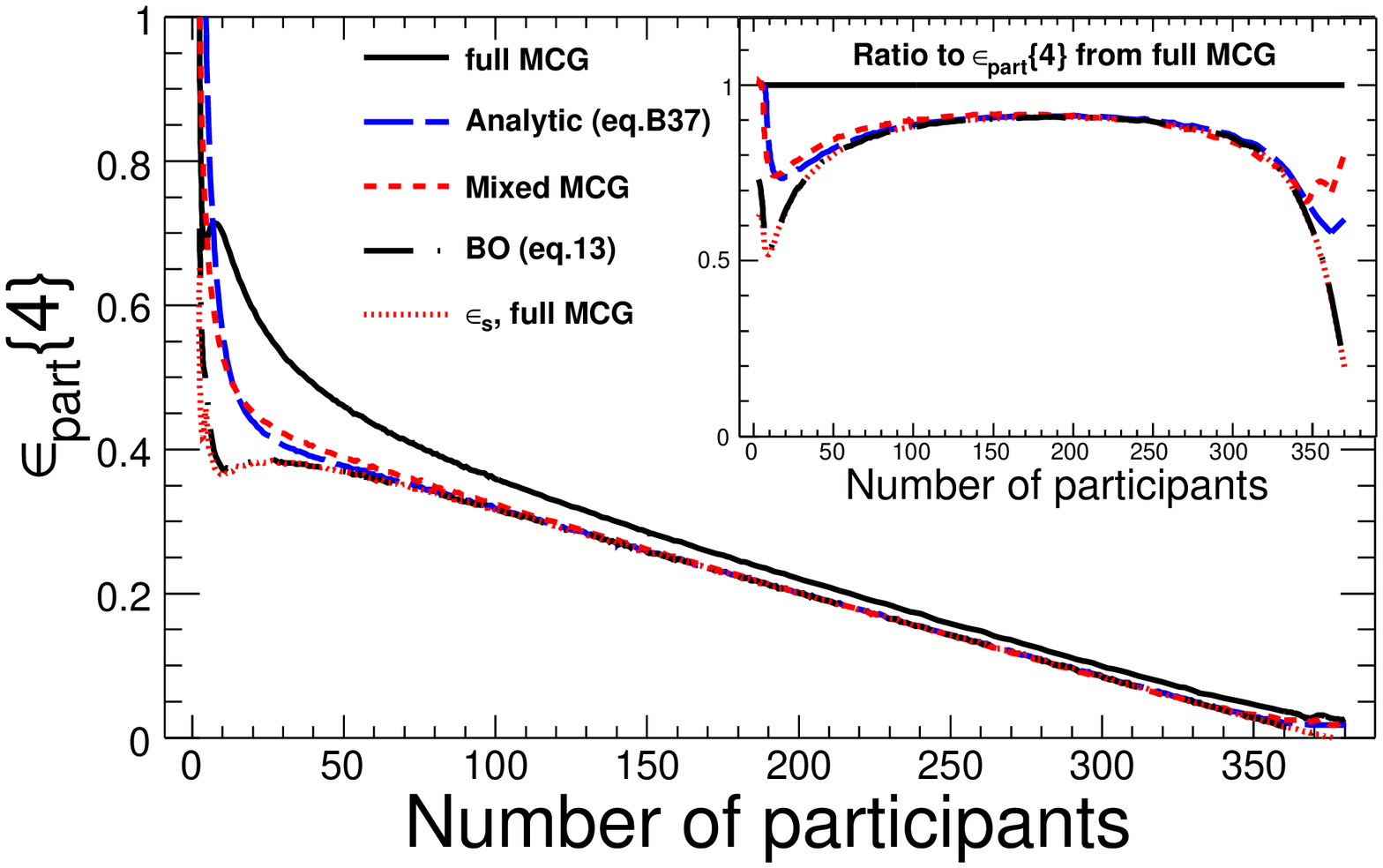}
\end{minipage}
\hspace{0.1cm}\hfill
\begin{minipage}[t]{0.49\textwidth}
\includegraphics[width=\textwidth]{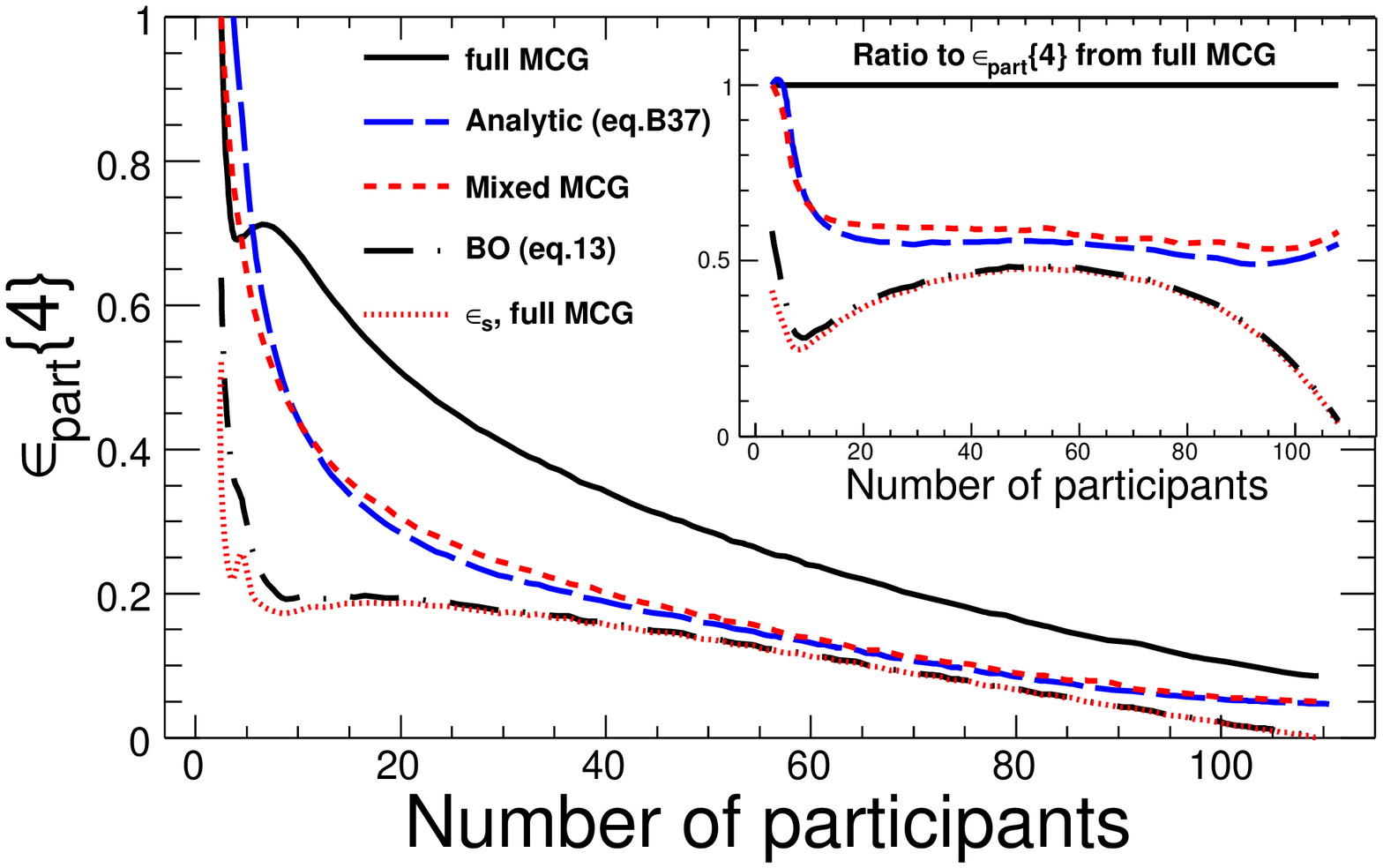}
\end{minipage}
\caption{\label{fig:ecc4comp}Comparison of $\epfour$ from the full \MCG~(upper set of curves), 
the mixed-event \MCG~(lower set of curves) with the semi-analytical approach, \Eq{eq:ecc4}, and
B\&O's approximation~(\Eqn{13} from \Ref{Bhalerao:2006tp}), as well as with the standard eccentricity 
as a function of $\Npart$ in \AuAu~(left) and \CuCu~(right) collisions at $\snn=200\,$GeV.
The inset shows the ratio \wrt\ $\epfour$ from the full \MCG\ calculation.}
\end{figure*}

\subsubsection{Cumulants in the Extended B\&O Approach}
The second approximation in \Ref{Bhalerao:2006tp} is to derive
expressions for $\eptwo$ and $\epfour$ using a Taylor expansion that
leads to a power series in $1/\Npart$. In \App{calccumulants}, we advance
the calculations to higher orders in $1/\Npart$ by generalizing \Eqn{11}
from \Ref{Bhalerao:2006tp}, and show how to obtain the analytical
terms in a rigorous fashion.  For $\eptwo$, we obtain the same
expression as \Eqn{12} in \Ref{Bhalerao:2006tp}, and prove that
the $\ordof{1/\Npart^2}$ terms are really negligible. We also
obtain the expansion for $\epfour$, \Eq{eq:ecc4}, where ---in contrast
to \Ref{Bhalerao:2006tp}--- all important terms have been kept. In
particular, for central collisions, when $\es \rightarrow 0$, some terms
of $\ordof{1/\Npart^3}$ are not negligible and must be kept. 

The values for the ensemble averages over participant nucleon 
distributions~(like for example $\av{r^2}$ or $\av{r^4\cos 2\phi}$) in 
\Eq{eq:ecc4} need to be calculated numerically. We calculate each of these
averages as a function of $\Npart$ using the usual~(full) PHOBOS \MCG\ code. 
Inserting the numerically evaluated values into \Eq{eq:ecc4}, leads to the 
``semi-analytic'' result discussed below.

\Fig{fig:ecc4comp} shows the results for $\epfour$, comparing the full PHOBOS 
and mixed-event \MCG\ with our semi-analytical result \Eq{eq:ecc4}, and with B\&O's 
semi-analytical approximation~(\Eqn{13} from \Ref{Bhalerao:2006tp} evaluated with 
the full PHOBOS \MCG), as well as with the standard eccentricity $\es$.
For both collision systems, our semi-analytical result fully agrees with 
the mixed-event \MCG\ calculation.
This is consistent with the fact that correlations among the participants 
are neglected in the analytical derivation of \Eq{eq:ecc4}. Furthermore, 
it confirms that all numerically important terms have been kept in \Eq{eq:ecc4}.
The full \MCG\ calculation which includes participant spatial correlations
disagrees with the other calculations that neglect them by almost a factor of two. 
Contrary to \Ref{Bhalerao:2006tp}, we find that, for the \CuCu\ system,
$\epfour$ calculated in the semi-analytical approach does not agree with $\es$,
in particular for very peripheral and near-central collisions.
More important, however, is the aforementioned effect of the neglected correlations. 
For the \AuAu\ system, $\epfour$ is found to be numerically close to $\es\,$(with
deviations of less than $10$\%) over a wide range of centralities. Only for very 
peripheral and near-central collisions correlations may play an important role.
For the \CuCu\ system, on the other hand, $\epfour$ differs from $\es$ by almost 
a factor of two over a wide range of centralities, implying that correlations
can not be neglected for the smaller system. 

\begin{figure*}[t]
\begin{minipage}[t]{0.49\textwidth}
\includegraphics[width=\textwidth]{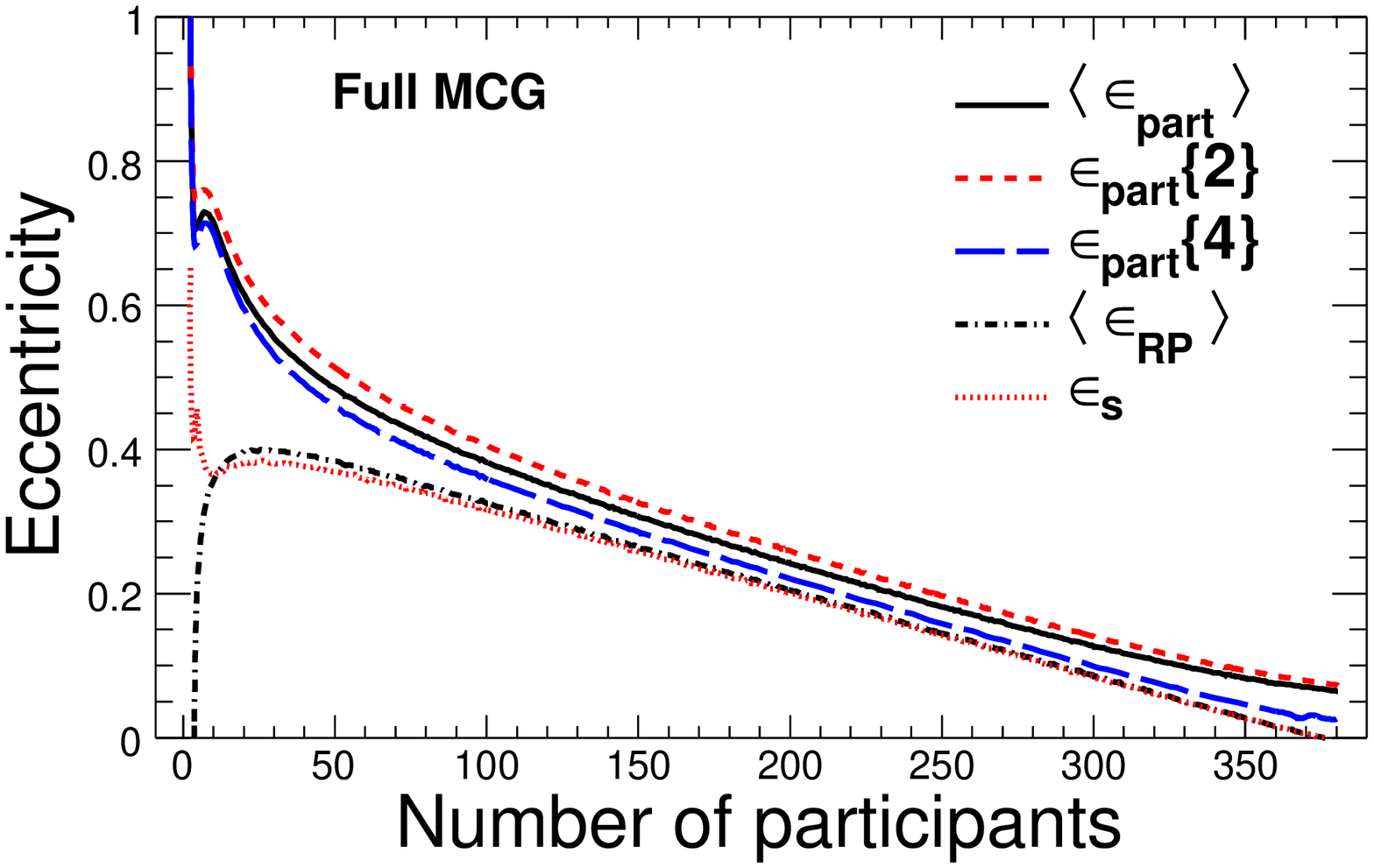}
\end{minipage}
\hspace{0.1cm}\hfill
\begin{minipage}[t]{0.49\textwidth}
\includegraphics[width=\textwidth]{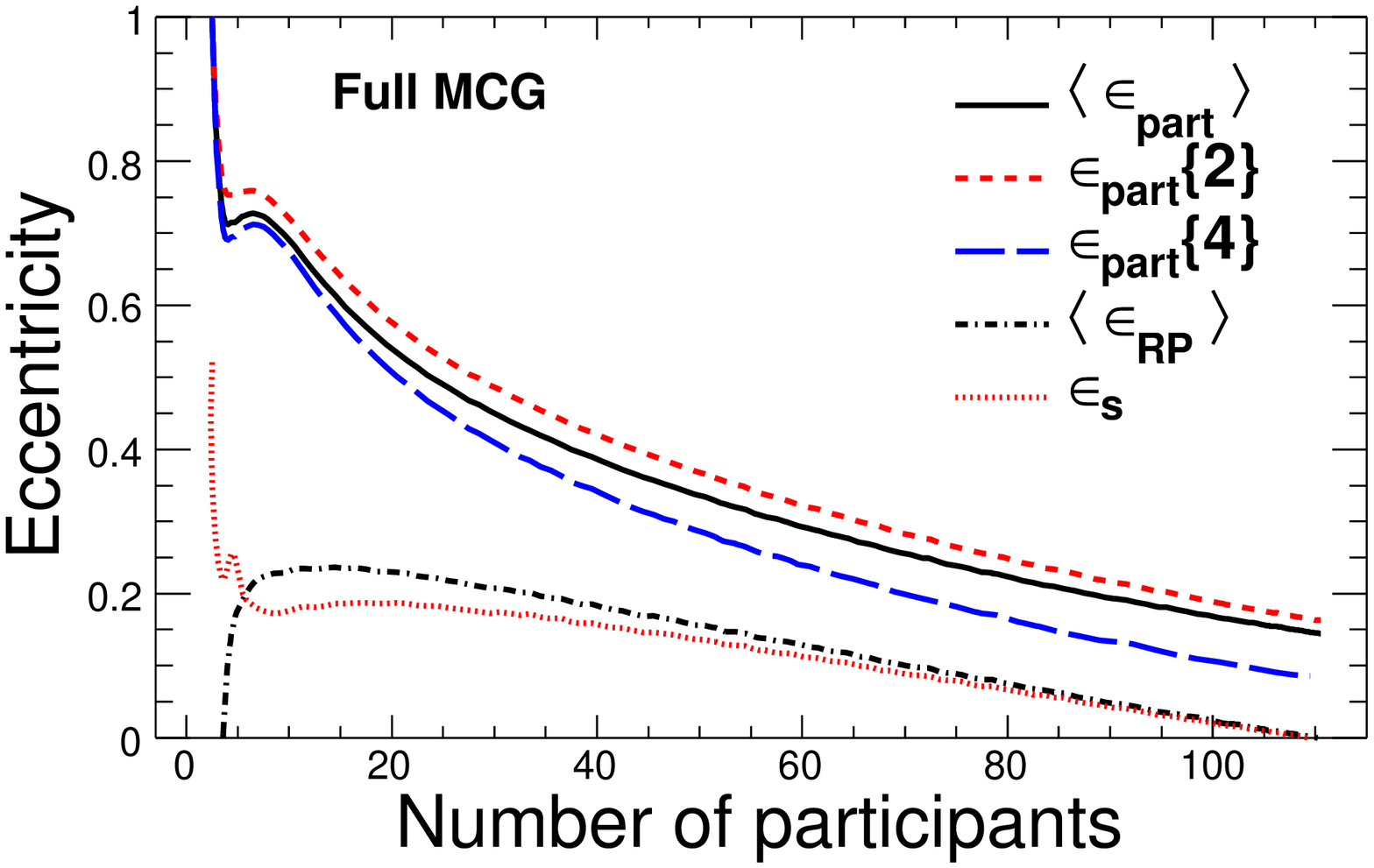}
\end{minipage}
\caption{\label{fig:eccfullmccomp}Participant eccentricity and cumulants, 
as well as reaction plane and standard eccentricities as a function of $\Npart$ 
in \AuAu~(left) and \CuCu~(right) collisions at $\snn=200\,$GeV obtained from 
full \MCG\ calculations.}
\end{figure*}

\Fig{fig:eccfullmccomp} shows a comparison of all eccentricity definitions
used in the present paper~(participant eccentricity and cumulants, reaction plane 
and standard eccentricity), obtained from full \MCG\ calculations with
baseline parameters listed in ~\Tab{tab:tab1}.

As mentioned before, in contrast to the results shown in \Fig{fig:ecc4comp}, 
the \mbox{authors} of \Ref{Bhalerao:2005mm} find in their work that
$\epfour$ differs very little from both the standard eccentricity
$\es$ and the average reaction plane eccentricity $\av{\erp}$, the
latter two being almost equal.
Recently, the authors of~\Ref{Voloshin:2007pc} 
have shown that, within a Gaussian model of the event-by-event eccentricity 
fluctuations, the identity of $\epfour$ with $\av{\erp}$ is exact~(see \Eqn{9} 
in~\cite{Voloshin:2007pc}) as long as the Gaussian widths for $\erp$ and for 
the correlation term $\rho_{xy} \equiv 2\sigma_{xy}/(\sigma_x^2 + \sigma_y^2)$ 
are equal.
We were able to trace the inequality between $\epfour$ and $\av{\erp}$ 
present in our \MCG\ model~(see \Fig{fig:eccfullmccomp}) to a breakdown of 
the Gaussian model assumptions made in \Ref{Voloshin:2007pc}, in particular for 
peripheral collisions and small collision systems. We find that in the \MCG\ model
the event-by-event fluctuations of the correlation term $\rho_{xy}$ are indeed Gaussian 
for mid-central to central \AuAu\ and most central \CuCu\ collisions. On the other hand, 
the event-by-event fluctuations of $\erp$ are not well described by a Gaussian function 
for all \AuAu\ and \CuCu\ collisions except the most central ones. Furthermore, for 
semiperipheral and peripheral collisions the width of the $\erp$ distribution does not 
agree with the width of $\rho_{xy}$. 
Consequently, for all but the most central collisions, 
our MCG model results are poorly described by the Bessel-Gaussian
distribution given in \Eqn{3} of \Ref{Voloshin:2007pc} on which
the equality of $\epfour$ and $\av{\erp}$ is based.

\section{\label{conclusions}Conclusions}

The interpretation of the anisotropic flow data measured in nucleus--nucleus collisions
at high energy requires a detailed understanding of the initial source anisotropy,
which is typically quantified by the eccentricity of the shape of the nuclear overlap
area. In this paper, we investigate various ways of defining this effective
eccentricity using \MCG\ calculations.

We find that variations in the Glauber parameters have only small effects
on the participant eccentricity for both the \AuAu\ and \CuCu\ collision systems, while 
the reaction plane eccentricity shows variations on the $10$\% level~(\Fig{fig:eccrob}). 
The generalization from participant-weighted to collision-weighted interaction point 
distributions leads to an increase in the obtained participant eccentricity, by a constant 
shift, similarly to what is known for the standard eccentricity~(\Fig{fig:bincoll}). 
Over a realistic range of parameters, the modeling of smeared matter distributions does 
not lead to significantly different results for the participant eccentricity~(\Fig{fig:smeared}). 
Thus, we conclude that reasonable variations in density parameters, the sources of matter, 
and their localization have only a small effect on the participant eccentricity. These
results support our initial idea~\cite{Manly:2005zy,Alver:2006wh} to use the participant 
eccentricity definition in conjunction with elliptic flow scaling. 

Depending on the event-plane resolution, fluctuations in the elliptic flow magnitude 
influence the measured ``mean'' $\vrp$, and for low resolution bias the measurement 
towards the RMS of the elliptic flow distribution. For a given event-plane resolution, 
we find that there is a simple connection with the appropriate moment of $v_2$,
which appears to be independent of the level of uncorrelated background~(\Fig{fig:alphares}). 
For the resolutions achieved with the PHOBOS elliptic flow method~\cite{Back:2002gz} this gives 
a scaling with the RMS of $v_2$ for all measured systems and energies.
We take this into account in the eccentricity-scaled elliptic flow~(\Fig{fig:scaledv2}) by 
presenting it as $\vrp/\eptwo$ together with individual systematic errors from data and 
MC parameter variations. The participant-eccentricity-scaled elliptic flow shows an almost linear 
scaling with the particle area density as predicted in \Refs{Heiselberg:1998es,Voloshin:1999gs} in 
the low density limit.

A rigorous attempt to analytically derive non-negligible contributions to cumulants of the
participant eccentricity distribution out to $\ordof{1/\Npart^3}$ confirms the expressions 
found in \Ref{Bhalerao:2006tp}, for all but $\epfour$, where our derivation,
\Eq{eq:ecc4}, for the first time keeps all leading order terms in the series.
The numerical evaluation of our analytical result for $\epfour$ 
agrees with $\epfour$ obtained with the mixed-event \MCG\ 
calculation, as expected, since both ignore all correlations among the participating 
nucleons. 
In comparison, the results obtained from full PHOBOS \MCG\ calculations, 
which include spatial correlations among the participants, 
imply that pairwise spatial correlations among the participants from the collision 
process itself are quite important, especially for the \CuCu\ system.
For \CuCu\, the contribution to the participant eccentricity cumulants is about $20$--$45$\%, 
while it is about $10$\% over most of the centrality range for \AuAu~(\Fig{fig:eccrat}). 
Furthermore, it turns out that $\epfour$ for the \CuCu\ system differs from $\es$ by about a
factor of two, while the difference for \AuAu\ is smaller and only about $10$\% for most of the 
centrality range~(\Fig{fig:ecc4comp}, \Fig{fig:eccfullmccomp}). 
Therefore, while correlations among participating nucleons may be neglected at the $10$\% accuracy
level over a wide range of centralities in the larger \AuAu\ system, they are crucially important
for the smaller \CuCu\ system and should not be neglected.
These results suggest that correlations among participants that are certainly present in nature 
and that are to a large extent implemented in full \MCG\ calculations are sufficiently important 
that they should be taken into account in any study where nuclear geometry is expected to play an 
important role.

\begin{acknowledgements}

Fruitful discussions with Arthur K.~Kerman, John W.~Negele and Jean-Yves Ollitrault
are acknowledged.

%
%
%
%
This work was partially supported by U.S.~DOE grants 
DE-AC02-98CH10886,
DE-FG02-93ER40802, 
DE-FG02-94ER40818,  
DE-FG02-94ER40865, 
DE-FG02-99ER41099, and
DE-AC02-06CH11357, by U.S.\ 
NSF grants 9603486, 
0072204,            
and 0245011,        
by Polish KBN grant 1-P03B-062-27(2004-2007),
by NSC of Taiwan Contract NSC 89-2112-M-008-024, and
by Hungarian OTKA grant (F 049823).
The work of U.H.\ was supported by the U.S.\ DOE under grant DE-FG02-01ER41190.
\end{acknowledgements}

\appendix
\section{\label{eccdef}Eccentricity Definitions}

The spatial anisotropy of the interaction region in the transverse plane~(in the following as throughout the paper given by 
the $x$- and $y$-axes) is commonly called ``eccentricity'' and denoted with the symbol $\myepsilon$. It has been introduced in 
\Ref{Heiselberg:1998es}~(called ``deformation'', symbol  $\delta$) and in \Ref{Sorge:1998mk}~(called ``spatial asymmetry'', 
symbol $\alpha_x$), while the basic idea of a dimensionless {\it momentum-based} anisotropy parameter~(symbol $\alpha$) originates 
from \Ref{Ollitrault:1992bk}. 
In its most basic formulation, the eccentricity definition reads
\beq
\myepsilon = \frac{R^2_y - R^2_x}{R^2_y+R^2_x}
\label{eq:epsilon}
\eeq
where $R^2_x$ and $R^2_y$ characterize the size of the source in the $x$- and $y$-direction, respectively. Note that this definition 
allows positive and negative values, $-1\le\myepsilon\le1$. It is related, but not identical, to the geometrical definition of 
the eccentricity of an ellipse. In this appendix, we will summarize the most important definitions of eccentricity.

\subsection{Standard Eccentricity}

Prior to our work~\cite{Manly:2005zy}, it has been common practise to use smooth, event-averaged initial conditions, for 
which the initial spatial asymmetry in the transverse plane has typically been given by the ``standard'' 
eccentricity
\beq
\es = \frac{\avsm{y^2}-\avsm{x^2}}{\avsm{y^2}+\avsm{x^2}}
\label{eq:epsstd}
\eeq
where $\avsm{x^2}$ and $\avsm{y^2}$ are the second moments of the (typically participant-weighted) ensemble-averaged nucleon distribution 
in the $x$- and $y$-direction, respectively.
We follow the notation introduced by Bhalerao \& Ollitrault in \Ref{Bhalerao:2006tp}, where $\avsm{\dots}$ denotes an average 
taken over many events~(ensemble average), while $\{\cdots\}$ stands for an average~(over participants)
in a single event~(sample average).

\subsection{Reaction Plane Eccentricity}

The two incoming nuclei, separated by the impact parameter $b$, can be assumed to be
centered at ($\pm b/2$,0) in the transverse plane such that for a given event the chosen 
MC frame coincides with the nuclear reaction frame. 
The ``reaction plane'' eccentricity is then obtained from
\beq
\erp = \frac{\sigma^2_y-\sigma^2_x}{\sigma^2_y+\sigma^2_x}
\label{eq:epsrp}
\eeq
where $\sigma^2_x = \{x^2\} -\{x\}^2$ and $\sigma^2_y = \{y^2\} -\{y\}^2$ are the (typically 
participant-weighted) variances of the nucleon distribution in $x$- and $y$-direction in a given event. 
In contrast to $\es$, which is only defined for the entire ensemble of collision events, $\erp$ is defined
for each event and has its own ensemble average $\av{\erp}$.
The reaction plane eccentricity can
be useful in comparisons with $v_2$ data where the reaction plane is determined by spectator neutrons in a zero-degree 
calorimeter~\cite{Bhalerao:2006tp}. 
Since $\av{\erp}$ is numerically very similar to $\es$ (in \AuAu\ the difference for $\Npart>10$ is at most 
$5$\%, however in \CuCu\ the difference is generally between $15$--$30$\%, see also \Fig{fig:eccfullmccomp}), 
it has been used in connection with \MCG\ calculations in place of the standard eccentricity~\footnote{In 
\Refs{Manly:2005zy,Alver:2006wh} the reaction plane eccentricity was instead named $\myepsilon_{\rm std}$ or 
$\myepsilon_{\rm standard}$.}.

\subsection{Participant Eccentricity}

The ``participant'' eccentricity expresses the overlap eccentricity in the rotated (``participant'') frame~(see 
\Fig{fig:eccpartscheme}), denoted by $x^\prime$ and $y^\prime$, which for a given event maximizes $\sigma^{\prime}_y$ 
and minimizes $\sigma^{\prime}_x$. In principle, the overlap zone will also be shifted \wrt\ the reaction plane frame, 
but this shift has no impact on the eccentricity. Generally, the second moments of the position distribution in the 
nuclear reaction plane (or MC) frame are described by the covariance matrix,
\beq
\Sigma = 
 \left( \begin{array}{cc}
 \sigma^2_x & \sigma_{xy} \\
 \sigma_{xy} & \sigma^2_y
 \end{array} \right )
\eeq
where $\sigma^2_x$, $\sigma^2_y$ and $\sigma_{xy} = \{xy\} - \{x\}\{y\}$ are the per-event (co-)variances of the underlying 
(typically participant weighted) nucleon distribution in the transverse plane, given in the original frame.
The participant frame corresponds to the frame in which $\Sigma$ is diagonal. Since $\Sigma$ is a real symmetric matrix, its 
diagonalization can be accomplished by finding the eigenvalues $\lambda$ that satisfy $\det(\Sigma - \lambda I) = 0$, leading 
to a second order polynomial in $\lambda$ with two solutions:
\beq
\lambda^{\pm} = 
     \frac{1}{2} \left( \sigma^2_y+\sigma^2_x \pm 
     \sqrt{(\sigma^2_y-\sigma^2_x)^2 + 4 \sigma_{xy}^2} \, \right)\,.
\eeq
These two values of $\lambda$ correspond to $\sigma^{\prime2}_x$ and $\sigma^{\prime2}_y$ with the larger 
value~($\lambda^{+}$) corresponding to the $y^\prime$ and the smaller value~($\lambda^{-}$) to the $x^\prime$
direction, by definition. This leads to the expression for the participant eccentricity~\cite{Manly:2005zy,Alver:2006wh},
\beq
\epart 
 = \frac{\sigma^{\prime2}_y-\sigma^{\prime2}_x}
        {\sigma^{\prime2}_y+\sigma^{\prime2}_x} 
 = \frac{ \sqrt{ (\sigma^2_y-\sigma^2_x)^2 + 4 \sigma_{xy}^2}}
        {\sigma^2_y+\sigma^2_x}\,.
\label{eq:epspart}
\eeq
Like $\erp$, the participant eccentricity is defined on an event-wise basis, however
in contrast to the previous definitions of the eccentricity, it is non-negative, 
covering the range $0 \le \epart \le 1$ by construction. The participant frame is tilted 
event-by-event by an angle of $\psipart$ \wrt\ the reaction plane, where 
\beq
\tan \psipart=\frac{\sigma_{xy}}{\sigma^2_{y}-\lambda^{-}} \,\,\, 
            (= \frac{\sigma_{xy}}{\lambda^{+}-\sigma^2_{y}})\,.
\eeq
It should also be noted that since the overlap ellipse is generally tilted, its area is not proportional to 
$\sigma_x \sigma_y$ as often assumed, but rather given by
\beq
S = \pi \, \sigma^\prime_x \sigma^\prime_y
  = \pi \, \sqrt{ \sigma^2_x \sigma^2_y - \sigma_{xy}^2}\,.
\label{eq:overlaps}
\eeq
Numerically the ratio, $\frac{\sigma^\prime_x\sigma^\prime_y}{\sigma_x \sigma_y}$, 
is very similar for the \CuCu\ and \AuAu\ systems at the same $\Npart$, larger than 0.75, 
and increasing with increasing centrality so that for $\Npart\ge 20$ it is larger than 0.9 
in $200\,$GeV collisions.

\section{\label{calccumulants}Calculating Cumulants}

In \Ref{Bhalerao:2006tp} Bhalerao \& Ollitrault~(B\&O) derive the behavior of 
various cumulant moments of the participant eccentricity analytically, making use of 
two questionable approximations. First, the paper contains an implicit assumption that 
all of the participant positions are independent samples of some underlying distribution, 
or at least that any correlations between participants are unimportant to the eccentricity
fluctuations~\cite{Ollitrault:2006privat}. Second, the paper uses a Taylor expansion
which leads to a power series in $1/\Npart$ which is then truncated at $\ordof{1/\Npart}$ 
without proof that the nominally higher order terms are actually smaller.

In this appendix, we will extend the B\&O results
for the case in which any correlations between participant positions are
still considered negligible, but keeping all important terms of the
Taylor expansion. In particular, we will generalize 
Eqs.~(11)--(14) of \Ref{Bhalerao:2006tp} and comment on \Eqn{8}.

\subsection{Generalizing B\&O Equation~11}

Following B\&O, we will assume that for each event $N\equiv \Npart$ participants
are generated independently from an arbitrary underlying 2-dimensional
distribution.  The averaging symbol $\av{f}$ denotes the average of
the quantity $f$ over the
underlying distribution and/or the ensemble average value taken over a large
number of events. In order to investigate fluctuations, we must also
consider event-wise averages $\{f\}\equiv
\frac{1}{N}\sum_{i=1}^{N}f_i$. The
event-by-event fluctuations are given by $\delta_f \equiv \{f\} -
\av{f}$. For convenience of calculation, let us also define $\hat{f}
\equiv f - \av{f}$ such that $\delta_f = \{\hat{f}\}$.

Obviously,
\beq
\av{\delta_f} = \avsm{\hat{f}} = 0\,.
\eeq

Next, we evaluate $\avsm{\delta_f \delta_g}$ by exhibiting the ensemble
and event averages explicitly:
\beqa
& & \av{\delta_f \delta_g} = \av{\{\hat{f}\}\{\hat{g}\}}
    = \frac{1}{N_{ev}N^2} \sum_{n=1}^{N_{ev}} \sum_{i=1}^N \hat{f}_{i,(n)}
      \sum_{j=1}^N \hat{g}_{j,(n)} \nonumber \\
& & = \frac{1}{N_{ev}N^2} \left(
      \sum_{i=1}^N \sum_{n=1}^{N_{ev}} \hat{f}_{i,(n)}\hat{g}_{i,(n)} + 
      \mathop{\sum_{i=1}^N}_{j\ne i} \sum_{n=1}^{N_{ev}} 
      \hat{f}_{i,(n)}\hat{g}_{j,(n)} \right) \nonumber \\
& & = \frac{1}{N^2} \left(
      \sum_{i=1}^N \av{\hat{f}_i\hat{g}_i} + 
      \mathop{\sum_{i=1}^N}_{j\ne i} \av{\hat{f}_i\hat{g}_j} 
      \right)\,.
\eeqa
The sums are all finite and their order can be
interchanged freely. If the participants are numbered randomly, 
then $\av{f_i}=\av{f}$. For example, the average of the
$x^2$ values for all ``participants number 7'' over all events will
just be $\av{x^2}$. Correspondingly,
$\avsm{\hat{f}_i}=\avsm{\hat{f}}=0$.
For $\avsm{\hat{f}_i\hat{g}_i}$ we have
\beq
 \avsm{\hat{f}_i\hat{g}_i}=\avsm{(\hat{f}\hat{g})_i}=\avsm{\hat{f}\hat{g}}\,.
\eeq
For $\avsm{\hat{f}_i\hat{g}_j}$ with $i\ne j$ we have
\beqa
\avsm{\hat{f}_i\hat{g}_j}=
\avsm{\hat{f}_i}\avsm{\hat{g}_j}
& = & \avsm{\hat{f}}\avsm{\hat{g}} = 0\,,
\label{eq:nocorr}
\eeqa
since, following B\&O, the positions of participants $i$ and $j\,$($i\neq j$) in each
event are assumed to be uncorrelated.
It is this last step in \Eq{eq:nocorr} which fails when there are correlations between
the locations of different participants. Neglecting such correlations, as in B\&O, 
one finally arrives at
\beqa 
\label{eg:delta2d}
\av{\delta_f \delta_g} & = & \frac{1}{N^2}
                             \left( N \avsm{\hat{f}\hat{g}} + 
                      N(N-1) \avsm{\hat{f}}\avsm{\hat{g}} \right) \nonumber  \\
& = & \frac{\avsm{\hat{f}\hat{g}}}{N} 
=  \frac{\av{(f-\av{f})(g-\av{g})}}{N} \nonumber \\
& = & \frac{\av{fg}-\av{f}\av{g}}{N} 
\eeqa
in agreement with \Eqn{11} from B\&O~\cite{Bhalerao:2006tp}.

Generalizing the above derivation to higher orders, we, as well as
Ollitrault~\cite{Ollitrault:2006privat}, find that the correct generalization
of \Eq{eg:delta2d} for $\delta^3$ terms is given by
\beq
\avsm{\delta_f \delta_g \delta_h} = \frac{\av{\hat{f}\hat{g}\hat{h}}}{N^2}\,.
\eeq
Compared to the $\delta^2$ terms, \Eq{eg:delta2d}, this is suppressed by a
factor $1/N$.
Starting with the $\delta^4$ term, the expressions begin to get more
complicated. In particular, 
\beqa
\avsm{\delta_f \delta_g \delta_h \delta_u} &=&
       \frac{\av{\hat{f}\hat{g}\hat{h}\hat{u}}}{N^3} + 
       \frac{(N-1)}{N^3} 
         \left( \avsm{\hat{f}\hat{g}}\avsm{\hat{h}\hat{u}} 
         \right. \nonumber \\
& &      + \left. \avsm{\hat{f}\hat{h}}\avsm{\hat{g}\hat{u}} + 
         \avsm{\hat{f}\hat{u}}\avsm{\hat{g}\hat{h}} 
         \right)\,,
\label{eq:delta4}
\eeqa
which is actually $\ordof{1/N^2}$, \ie~the same order in the number
of participants as the $\delta^3$ term.

The fifth and sixth order terms can be calculated similarly. Any terms
involving single powers like $\avsm{\hat{f}}$ will again vanish. 
The nonzero terms are
\beqa
\av{\delta_f \delta_g \delta_h \delta_u \delta_v} 
& = & \frac{1}{N^4} \av{\hat{f}\hat{g}\hat{h}\hat{u}\hat{v}}+ 
      \frac{N-1}{N^4} 
            \left( \avsm{\hat{f}\hat{g}\hat{h}}\avsm{\hat{u}\hat{v}} 
            \right. \nonumber \\
& & + \,
                   \avsm{\hat{f}\hat{g}\hat{u}}\avsm{\hat{h}\hat{v}} + 
                   \avsm{\hat{f}\hat{g}\hat{v}}\avsm{\hat{h}\hat{u}} +
 	           \avsm{\hat{f}\hat{h}\hat{u}}\avsm{\hat{g}\hat{v}}  
            \nonumber \\
& & + \,
                   \avsm{\hat{f}\hat{h}\hat{v}}\avsm{\hat{g}\hat{u}} + 
                   \avsm{\hat{f}\hat{u}\hat{v}}\avsm{\hat{g}\hat{h}} + 
                   \avsm{\hat{g}\hat{h}\hat{u}}\avsm{\hat{f}\hat{v}}  
	    \nonumber \\
& & +       \left.
                   \avsm{\hat{g}\hat{h}\hat{v}}\avsm{\hat{f}\hat{u}} + 
                   \avsm{\hat{g}\hat{u}\hat{v}}\avsm{\hat{f}\hat{h}} + 
                   \avsm{\hat{h}\hat{u}\hat{v}}\avsm{\hat{f}\hat{g}} 
            \right) \nonumber \\
\label{eq:delta5}
\eeqa
and
\beqa
\av{\delta_f \delta_g \delta_h \delta_u \delta_v \delta_w} 
& = &  \frac{1}{N^5} \left[ \av{\hat{f}\hat{g}\hat{h}
                           \hat{u}\hat{v}\hat{w}}  \right.\nonumber \\ 
& &     +\, (N-1) \left( \avsm{\hat{f}\hat{g}\hat{h}\hat{u}}
                         \avsm{\hat{v}\hat{w}} + {\rm 14\ perms.} 
          \right)  \nonumber \\
& &     +\, (N-1) \left( \avsm{\hat{f}\hat{g}\hat{h}}
                         \avsm{\hat{u}\hat{v}\hat{w}} + {\rm 9\ perms.} 
          \right)  \nonumber \\
& &     +\, (N-1)(N-2) \times \nonumber \\ 
& &     \left. \,\,\, \left(\avsm{\hat{f}\hat{g}}\avsm{\hat{h}\hat{u}}
               \avsm{\hat{v}\hat{w}} + {\rm 14\ perms.} 
           \right) \right].
\label{eq:delta6}
\eeqa

In general, the dominant terms in the $1/N$ expansion should be those
composed of products of bilinears like $\av{\hat f\hat g}$ in the case of
even powers of $\delta$, and those with bilinears and one trilinear in the 
case of odd powers of~$\delta$. So, we have 
$\ordof{\delta^{2n-1}} = \ordof{1/N^n}$ for $n  > 1$ and
$\ordof{\delta^{2n  }} = \ordof{1/N^n}$ for $n\ge 1$. 
This means that we must consider terms up to $\ordof{\delta^4}$ if we
want to find all $\ordof{1/N^2}$ terms in the series truncated by
B\&O, and terms up to $\ordof{\delta^6}$ in order to
capture the leading behavior in the limit $\es \rightarrow 0\,$(see below).

\subsection{B\&O Equation~8}

The Taylor expansion which leads to \Eqn{8} in B\&O is not applicable everywhere
as it implicitly assumes that $1/(N \es^2) \ll 1$.
Since $\es\rightarrow 0$ for central collisions, this quantity is not
guaranteed to be small, and this expansion is poorly behaved and formally
divergent for central collisions.  Fortunately, \Eqn{8} of B\&O is
not actually needed in order to derive and generalize their
\mbox{Eqs.~(12)--(14)}.

\subsection{B\&O Equation~12: Calculating $\eptwo^2$}

In order to calculate $\eptwo^2=\av{\epsq}$, we must first express 
$\epsq$ in terms of the $\delta$'s. 
We start with the definition, \Eq{eq:epspart}: 
\beq
\epsq = \frac{(\sigma^2_y-\sigma^2_x)^2+4\sigma^2_{xy}}{(\sigma^2_y+\sigma^2_x)^2}\,.
\eeq
Following B\&O we have
\beqa
\sigma^2_{x\hide{y}} & = & \{x^2\} - \{x\}^2 = \av{x^2} + \delta_{x^2} - \delta^2_{x} \\
\sigma^2_{y\hide{y}} & = & \{y^2\} - \{y\}^2 = \av{y^2} + \delta_{y^2} - \delta^2_{y} \\
\sigma_{xy} & = & \{xy\}-\{x\}\{y\} = \delta_{xy}-\delta_x\delta_y
\eeqa
using $\av{x}=\av{y}=\av{xy}=\av{xy^n}=\av{x^ny}=0$.
This leads to the exact result:
\beqa
\epsq & = & 
  \left[ \es^2 + \boden{2\es\dysqm}{} 
         + \boden{\dysqm^2}{2} + \boden{4\delta^2_{xy}}{2}
  \right. \nonumber \\ 
& & 
         - \, \boden{2\es\dsqym}{} - \boden{2\dysqm\dsqym}{2} 
          \nonumber \\ 
& & \left .
         - \, \boden{8\delta_{xy}\delta_x\delta_y}{2}
         + \boden{\dsqym^2}{2} 
         + \boden{4\delta^2_x\delta^2_y}{2} \right] \times
\nonumber \\ 
& & 
\left[1 + \boden{\dysqp}{} -\boden{\dsqyp}{} \right]^{-2}\,,
\label{eq:ep2exact}
\eeqa
where $\av{r^2}=\av{x^2}+\av{y^2}$.
The second and third terms in the denominator are genuinely $\ll 1$,
so it can be safely Taylor expanded. The resulting polynomial 
series is well behaved.
The leading terms are $\es^2$ and $\ordof{\es^0/N}$. 
All terms of $\ordof{\delta^3}$ and higher are $\ordof{\es^0/N^2}$ and can 
be dropped since they are at least a full power of $1/N$ down without any 
compensating $1/\es$ factors. So, we obtain
\beqa
\epsq & = & 
    \es^2 + \boden{2\es\dysqm}{} -\boden{2\es^2\dysqp}{}
    + \boden{\dysqm^2}{2} + \boden{4\delta^2_{xy}}{2} \nonumber \\ 
& & -\, \boden{2\es\dsqym}{} - \boden{4\es\dsqysqm}{2}
    + \boden{2\es^2\dsqyp}{} \nonumber \\
& & +\, \boden{3\es^2\dysqp^2}{2} 
    + \ordof{\es^0\delta^3} +\ordof{\es^0\delta^4} \nonumber \\
& & +\, \ordof{\es^0\delta^5} +\ordof{\es^0\delta^6} + \ldots\,.
\eeqa
This leads to the same result as B\&O \Eqn{12}, except that we have
also shown that all further terms are subdominant:
\beqa
\av{\epsq} & = & \es^2 
    + \frac{1}{N\rsq^2} \left[ (1+3\es^2) \av{r^4} + 4\es\av{r^4\cos2\phi} \right]
  \nonumber \\
& & + \,\ordof{\frac{\es^0}{N^2}}
    + \ldots \,.
\eeqa
Similarly, \Eqn{14} in B\&O is well-behaved and correctly contains all of the leading 
terms. As noted above, this is different for the expansion of \Eqn{8} in B\&O, which
does not converge in the limit $\es \rightarrow 0$.

\subsection{B\&O Equation~13: Calculating $\epfour^4$}

We know from B\&O that the $\ordof{\es^2/N}$ terms cancel, leaving
the $\es^4$ term as apparently dominant. However, in order to confirm this,
we need to check that all of the nominally higher order terms are
actually small. 

In order to organize the calculation, let us write the expansion of
\Equa{eq:ep2exact} as:
\beq
\epsq = \es^2 + {\cal A + B + C + D + \ldots}
\label{eq:ep2skeleton}
\eeq
where ${\cal A}$ contains all terms of $\ordof{\delta}$, ${\cal B}$ of
$\ordof{\delta^2}$, and so on. Furthermore, let us define 
${\cal B}_0=\lim_{\es\rightarrow 0} {\cal B}$ and 
${\cal C}_0=\lim_{\es\rightarrow 0} {\cal C}$ 
\etc, so that
\beq
\lim_{\es\rightarrow 0} \epsq = {{\cal B}_0 + {\cal C}_0 + {\cal D}_0}\,.
\eeq
Explicitly, we will need the following equations:
\beqa
{\cal A} & \equiv & \boden{2\es\dysqm}{} - \boden{2\es^2\dysqp}{} 
 \\
{\cal B} & \equiv & \boden{\dysqm^2 + 4\delta^2_{xy}}{2}
  - \boden{2\es\dsqym}{} - \boden{4\es\dsqysqm}{2} \nonumber \\
&& +\, \boden{2\es^2\dsqyp}{} + \boden{3\es^2\dysqp^2}{2} 
 \\
{\cal C} & \equiv & 
  -\boden{2\dysqm\dsqym}{2} - \frac{8\delta_{xy}\delta_x\delta_y}{\rsq^2}
  -\boden{2\dysqm^2\dysqp}{3} \nonumber \\ 
 & & 
  -\, \boden{8\delta^2_{xy}\dysqp}{3}
  +\boden{4\es\dysqp\dsqym}{2} 
  +\boden{4\es\dysqm\dsqyp}{2}  \nonumber \\
 & & 
  +\, \boden{6\es\delta^2_{r^2}\dysqm}{3}
  -\boden{6\es^2\dysqp\dsqyp}{2} - \boden{4\es^2\dysqp^3}{3} \nonumber \\
 \\
{\cal B}_0 & \equiv & \boden{\dysqm^2+4\delta^2_{xy}}{2} 
 \\
{\cal C}_0 & \equiv & 
  -\boden{2\dysqm\dsqym}{2} 
  -\frac{8\delta_{xy}\delta_x\delta_y}{\rsq^2} \nonumber \\
& &
  -\, \boden{2(\dysqm^2+4\delta^2_{xy})\dysqp}{3}
 \\ 
{\cal D}_0 & \equiv & 
  \boden{\dsqyp^2}{2} +\boden{4\dsqysqm\dsqym}{3} \nonumber \\ 
& &
  +\, \boden{16\delta_{xy}\delta_x\delta_y\dysqp}{3} 
  +\boden{2(\dysqm^2+4\delta^2_{xy})\dsqyp}{3} \nonumber \\
& &
 +\,\boden{3(\dysqm^2+4\delta^2_{xy})\dysqp^2}{4}\,.
\eeqa
We can now calculate $\epfour^4$:
\beqa
\epfour^4 & \equiv & 2\av{\epsq}^2-\av{\epfr} \nonumber \\
& = & \es^4 - \av{{\cal A}^2} + 2\es^2\av{{\cal B}} 
  + 2\es^2\av{{\cal C}_0} - 2\av{\cal{AB}} \nonumber \\
& & +\, 2\es^2\av{{\cal D}_0} - 2\av{{\cal AC}}
  + 2\av{\cal B}^2 - \av{{\cal B}^2} \nonumber \\
& & +\, 4\av{{\cal B}_0}\av{{\cal C}_0} 
  - 2\av{{\cal B}_0 {\cal C}_0} 
  + 4\av{{\cal B}_0}\av{{\cal D}_0} \nonumber \\
& & -\, 2\av{{\cal B}_0 {\cal D}_0} -\av{{\cal C}_0^2}
\label{eq:epart4sq}
\eeqa
where we have kept all terms up to $\ordof{1/N}$, and the leading
terms in $\es$ at $\ordof{1/N^2}$ and $\ordof{1/N^3}$.
We evaluate the expressions in \Eq{eq:epart4sq} piece by piece, dropping any terms
that would contribute to $\epfour^4$ at $\ordof{\es^4/N^2}$ or
$\ordof{\es^n/N^3}$ for each $n>0$. 
Note that $\av{r^m\cos 2n\phi}=\ordof{\es^n}$.
This leads to the following expressions:
\beqa
\av{\cal B} & = & \frac{1}{N\rsq^2} 
  \left[(1+3\es^2)\av{r^4} + 4\es\av{r^4\cos 2\phi}\right] \nonumber \\
 \\
\av{{\cal B}_0} & = & \frac{\av{r^4}}{N\rsq^2} 
 \\
\av{{\cal A}^2} & = & \frac{1}{N\rsq^2} 
  \left[(2\es^2+4\es^4)\av{r^4} + 8\es^3\av{r^4\cos 2\phi} \right. \nonumber \\
& & \hspace{1.25cm}\left. + 2\es^2\av{r^4\cos 4\phi}\right] 
 \\
\av{{\cal C}_0} & = & \frac{1}{N^2} 
        \left[-\frac{2\av{r^6}}{\rsq^3} \right] 
 \\
\av{\cal AB} & = & \frac{1}{N^2} \left[
    -6\es^2\frac{\av{r^6}}{\rsq^3} -2\es\frac{\av{r^6\cos 2\phi}}{\rsq^3}
    \right]
 \\
\av{{\cal D}_0} & = & \frac{1}{N^2} 
    \left[ 2 - \frac{\av{r^4}}{\rsq^2} + \frac{3\av{r^4}^2}{\rsq^4} 
    \right] 
 \\
\av{{\cal AC}} & = & \frac{1}{N^2\rsq^4} 
    \left[ 10\es^2\av{r^4}^2 + 8\es\av{r^4}\av{r^4\cos 2\phi} \right] \nonumber \\
 \\
\av{\cal B}^2 & = & \frac{1}{N^2\rsq^4} 
    \left[(1+6\es^2)\av{r^4}^2 + 8\es\av{r^4}\av{r^4\cos 2\phi}\right] \nonumber \\
 \\
\av{{\cal B}^2} & = & \frac{1}{N^2} \left[
    4\es^2 - 2\es^2\frac{\av{r^4}}{\rsq^2} 
    +(2+14\es^2) \frac{\av{r^4}^2}{\rsq^4} \right. \nonumber \\
& & \hspace{0.7cm} \left. + 16\es\frac{\av{r^4}\av{r^4\cos 2\phi}}{\rsq^4} \right] 
+ \frac{\av{r^8}-2\av{r^4}^2}{N^3\rsq^4} \nonumber \\
\av{{\cal B}_0 {\cal C}_0} & = & \frac{1}{N^3}\left[
\frac{4\av{r^4}^2}{\rsq^4}
    -\frac{8\av{r^4}\av{r^6}}{\rsq^5} \right] 
 \\
\av{{\cal B}_0 {\cal D}_0} & = & \frac{1}{N^3} \left[
  \boden{2\av{r^4}}{2} -\boden{2\av{r^4}^2}{4} + \boden{6\av{r^4}^3}{6} 
   \right]
 \\
\av{{\cal C}_0^2} & = & \frac{1}{N^3} \left[
  \boden{4\av{r^4}}{2} -\boden{8\av{r^4}^2}{4} + \boden{8\av{r^4}^3}{6} 
   \right]\,.
\eeqa

\newpage
Assembling, this leads to the final result

\beqa
\epfour^4 & = &
\es^4 + \frac{1}{N\rsq^2}[2\es^4\av{r^4}-2\es^2\av{r^4\cos4\phi}] \nonumber \\ 
& & +\, \frac{1}{N^2} \left[ 8\es^2\frac{\av{r^6}}{\rsq^3} 
    + 4\es\frac{\av{r^6\cos2\phi}}{\rsq^3} \right. \nonumber \\
& & \left. \;\;\;\;\;\;\;\;\;\; -\, 16\es^2\frac{\av{r^4}^2}{\rsq^4}
    -16\es\frac{\av{r^4}\av{r^4\cos2\phi}}{\rsq^4} \right]  \nonumber \\
& & +\, \frac{1}{N^3} \left[ \frac{2\av{r^4}^2}{\rsq^4} -\frac{\av{r^8}}{\rsq^4}
    +\frac{8\av{r^4}\av{r^6}}{\rsq^5} \right. \nonumber \\
& & \left. \;\;\;\;\;\;\;\;\;\; -\, \frac{8\av{r^4}^3}{\rsq^6} \right] 
    + \ordof{\frac{\es^4}{N^2}} + \ordof{\frac{\es^2}{N^3}} \nonumber \\
& & +\,\ordof{\frac{\es^0}{N^4}} + \ordof{\frac{\es^0}{N^5}} + \ldots\,,
\label{eq:ecc4}
\eeqa
where we now have all of the leading terms. Terms which have been
dropped are down from the leading terms by at least a full factor of
$1/N$ without any compensating $1/\es$ factor. B\&O~\cite{Bhalerao:2006tp} 
left out the $\ordof{1/N^2}$ term and most importantly the $\ordof{1/N^3}$ term. 
For \CuCu, the $\ordof{1/N^3}$ term tends to be comparable to the ``leading'' 
$\es^4$ term.  For central collisions, as $\es$ vanishes, the $\ordof{1/N^3}$ 
term becomes dominant and certainly cannot be neglected. 

\bibliographystyle{apsrev}
\bibliography{ecc_paper}

\end{document}